\documentclass[aps,prc,twocolumn,showpreprintnumbers,floatfix,nofootinbib]{revtex4-2}
\usepackage{amsmath,amssymb,graphicx,bm}
\usepackage{bbm}
\usepackage{color}
\usepackage{hyperref}
\usepackage[normalem]{ulem}
\usepackage{slashed}

\allowdisplaybreaks[4] 
\newcommand{\vect}[1]{\bm{\mathrm{#1}}}
\newcommand{\pv}{\vect{p}}
\newcommand{\qv}{\vect{q}}
\newcommand{\sigmav}{\vect{\sigma}}
\newcommand{\sv}{\vect{s}}

\newcommand{\nablav}{\vect{\nabla}}
\newcommand{\laplacian}{\Delta}
\newcommand{\tauv}{b}
\newcommand{\nubar}{\overline{\nu}}

\DeclareMathOperator{\Imag}{Im}
\renewcommand{\Im}{\Imag}
\newcommand{\ncdot}{\!\cdot\!}
\newcommand{\ntimes}{\!\!\times\!}
\renewcommand{\emph}[1]{\textit{#1}}

\newcommand{\tmin}{\text{min}}
\newcommand{\tmax}{\text{max}}
\newcommand{\produ}{\text{produ}}
\newcommand{\ph}{\text{ph}}
\newcommand{\RPA}{\text{RPA}}
\newcommand{\Landau}{\text{Landau}}
\newcommand{\Sign}{\mathbb{S}}

\begin{document}
\title{Charged current neutrino processes in hot nuclear matter with a recent Skyrme parametrization constrained by microscopic calculations}

\author{Mingya Duan} 
\affiliation{Universit\'e Paris-Saclay, CNRS/IN2P3, IJCLab, 91405 Orsay, France}
\author{Michael Urban} 
\email{michael.urban@ijclab.in2p3.fr}
\affiliation{Universit\'e Paris-Saclay, CNRS/IN2P3, IJCLab, 91405 Orsay, France}

\begin{abstract}
Neutrino processes are important in the modeling of supernova explosions, proto-neutron star evolution, and binary neutron star mergers.
We study neutrino production and absorption in proto-neutron star and supernova matter and direct Urca neutrino emission of neutron star matter in the framework of the random phase approximation (RPA).
As interactions, we employ the recent extended Skyrme parametrization Sky3s whose effective masses and spin-dependent terms were adjusted to microscopic calculations, and the SLy4 parametrization that was used in previous calculations of neutrino rates.
The rates obtained for Sky3s differ from those for SLy4 by up to one order of magnitude for some processes and energy regions.
We also determine the electron, muon, and proton fractions that lead to a stationary composition of matter for a density above the direct Urca threshold, and find that with Sky3s the standard $\beta$ equilibrium condition is not as badly violated at finite temperature as predicted in the literature.
There are also minor differences between the full RPA and the common Landau approximation, but they are probably not significant for astrophysical simulations.
We conclude that it would be worthwhile to repeat the calculation of neutrino rates for the use in astrophysical simulations, and the corresponding simulations, with several and better constrained interactions than SLy4, such as Sky3s.
\end{abstract}

\maketitle

\section{Introduction}\label{sec:introduction}
During the core collapse of a massive star ($M \gtrsim 8 M_{\odot}$,
$M_{\odot}$ is the solar mass), the core can reach densities
comparable with nuclear saturation density ($\rho_0 \approx 0.16$
nucleons/fm$^3$). During and after the collapse, a large number of
neutrons and neutrinos of all flavours are produced.  At high enough
temperatures, they can interact frequently with nuclear matter before
they escape from the star.  Neutrino production, absorption, and
scattering rates are essential ingredients in the modeling of
supernovae, neutron stars, and binary neutron star mergers.  In a
pioneering work, Bruenn incorporated neutrino transport into the
simulation \cite{Bruenn1985}.  Since the 2000s, the treatment of
neutrino transport in simulations has been continuously improved
\cite{Rampp2000,Mezzacappa2001,Rampp2002,Thompson2003,Buras2006,Horowitz2006}.
Many studies have also aimed to improve the treatment of
neutrino-matter interaction, such as taking into account relativistic
kinematics
\cite{Roberts2017,Fischer2020a,Guo2020,Sugiura2022,Sokolowski2026},
considering the role of muons
\cite{Bollig2017,Guo2020,Fischer2020b,Sugiura2022,Loffredo2023,Gieg2025,Sokolowski2026},
employing chiral effective field theory (EFT)
\cite{Bacca2009,Bacca2012,Rrapaj2015,Guo2019,Vidana2022,Shin2024},
using the Brueckner-Hartree-Fock (BHF) formalism with the two-body
potential of chiral EFT \cite{Guo2024}, and so on.

From a theoretical point of view, the neutrino rates (production, absorption, and scattering) can be directly related to the current-current correlation functions (response functions).
Neutrino scattering can be related to the neutral-current response functions, while the neutrino production and absorption can be related to the charged-current response functions.
The response functions depend on the nuclear interaction.
Among the different nuclear interactions, Skyrme effective interactions are easy to use due to their zero-range character.
Especially, the solution of the full random phase approximation (RPA) with Skyrme interactions is relatively easy.
In the late 20th century, Skyrme functionals had already been used to study neutrino interactions in hot and dense matter at the level of the Landau approximation \cite{Reddy1998,Reddy1999}, including the neutral- and charged-current interactions.
Concerning the charged-current interactions with Skyrme interactions, Ref. \cite{Dzhioev2018} studied the electron neutrino and antineutrino absorptions ($\nu_e + n \rightarrow p + e^{-}$ and $\bar{\nu}_e + p \rightarrow n + e^{+}$) in hot and dense supernova matter using the full Skyrme RPA in 2018.
Very recently, improved charged-current neutrino-nucleon interactions in dense and hot matter were studied \cite{Oertel2020} and applied to numerical simulations of proto-neutron star evolution \cite{Pascal2022}, using the Skyrme functional SLy4 \cite{Chabanat1998}. 

However, there are still some open problems.
On the one hand, in the above Refs. \cite{Oertel2020,Pascal2022}, the neutrino rates were only computed at the level of the Landau approximation instead of the full RPA.
On the other hand, many Skyrme interactions predict an unrealistic density dependence of the effective masses, resulting in a neutron Fermi velocity that is larger than the speed of light already at relatively low densities \citep{Duan2023}.
The widely used SLy4 parametrization is not an exception.
Besides, like other Skyrme functionals, SLy4 does not have well-constrained spin-dependent terms \cite{Duan2025}. 

To solve the above problems of previous Skyrme functionals, we
recently constructed two alternative extended Skyrme interactions,
called Sky3s and Sky4s. Their nucleon effective masses were fitted in
Ref. \cite{Duan2024} to have a more realistic density dependence as
predicted by microscopic BHF calculations employing the Argonne V18
interaction plus three-body force (TBF) \cite{Baldo2014} (which agree
with those of Ref. \cite{Shang2020} at zero temperature), solving at
the same time the problem of the neutron Fermi velocity mentioned
above. The other parameters that determine the mean field potentials
were, in addition to the fit of masses and radii of doubly-magic
nuclei, adjusted to microscopic EoS calculations based on chiral
interactions for pure neutron matter at the lowest densities
\cite{Palaniappan2023} and to BHF calculations with TBF at
intermediate and high densities \cite{Liu2022} for Sky3s, while the
somewhat softer EoS from Ref. \cite{Akmal1998} was used for Sky4s.
Furthermore, in Ref. \cite{Duan2025}, the spin-dependent terms of
Sky3s and Sky4s were determined by fitting the Landau parameters $G_0$
and $G_0^{\prime}$ in neutron matter and symmetric nuclear matter and
the effective-mass splitting of up and down particles in
spin-polarized matter to the results of microscopic calculations
\cite{Zuo2003SNM,Zuo2003communications,Zuo2003PNM,Vidana2016,Bigdeli2009}.
Therefore, to obtain high-quality neutrino transport for the
simulations of supernova explosions, proto-neutron star evolution, and
binary neutron star mergers, it is useful to develop and study the
charged-current response functions and the corresponding neutrino
rates with these new Skyrme parametrizations. It is worth
  mentioning that, unlike in microscopic calculations such as
  Ref. \cite{Shang2020}, the nucleon effective masses are temperature
  independent in our Skyrme parametrizations.

In this paper, we first summarize in Sec. \ref{sec:formalism} the derivation of the production and absorption rates and the computation of the charged-current RPA response functions with extended Skyrme interactions in asymmetric nuclear matter.
Then we present the charged-current response functions and the corresponding neutrino rates at different levels of approximation, computed using the new Skyrme parametrization Sky3s in Sec. \ref{sec:results} and compare them with the corresponding SLy4 results.
Finally, we summarize and conclude in Sec. \ref{sec:conclusion}.
For completeness, the expressions needed for the RPA are given in the appendix.

\section{Formalism}
\label{sec:formalism}
\subsection{Charged current neutrino interactions in hot nuclear matter} 
\label{subsec:charged current interactions in hot nuclear matter}
Although $\tau$ leptons are not produced because of their high mass, as a consequence of neutrino flavor oscillations, neutrinos of all flavours can be produced during a core-collapse supernova explosion \cite{Volpe2024}.
Among the three flavours of neutrinos, the $\mu$ and $\tau$ neutrino charged current absorption reactions are generally kinematically suppressed \citep{Reddy1998}.
The charged-current processes
\begin{align}\label{eq:charged-current processes1}
& e^{-} + p \leftrightarrow \nu_e + n,  \qquad e^{+} + n \leftrightarrow \nubar_e + p,
\end{align}
are the most important neutrino processes in supernova and proto-neutron star matter \citep{Janka2017a}. They were already included in the first relevant simulations \citep{Bruenn1985}.
Recently, the following charged-current processes
\begin{align}\label{eq:charged-current processes2}
p \leftrightarrow n+e^{+}+\nu_e,  \qquad n \leftrightarrow p+e^{-} + \nubar_e, 
\end{align}
have also been considered in simulations of supernova explosion and proto-neutron star evolution \citep{Oertel2020,Pascal2022,Fischer2020a}.

As the neutrino scattering rates can be related to the neutral-current response functions, the neutrino production and absorption rates can be related to the charged-current response functions.
In this section, we recall the derivation of the relation between the (anti)neutrino production and absorption rates and the nuclear density and spin-density correlation functions using Fermi's golden rule.
Throughout this paper, we use units with $\hbar = c = k_B = 1$, where $\hbar$, $c$, and $k_B$ are the reduced Planck constant, the speed of light, and the Boltzmann constant, respectively.

First, let us consider a general charged-current process
\begin{equation}\label{eq:general charged-current process}
l + b \rightarrow l^{\prime} + b^{\prime},
\end{equation}
which includes an incoming lepton $l$, an incoming baryon (nucleon) $b$, an outgoing lepton $l^{\prime}$, and an outgoing baryon (nucleon) $b^{\prime}$. In this process, one of the leptons is a neutrino (antineutrino).
Similar to the scattering between neutrinos and nucleons \citep{Iwamoto1982}, the charged-current interaction Lagrangian density is \citep{Reddy1998}
\begin{equation}\label{eq:lagrangian density}
\mathcal{L}_{I}(x) = \frac{G}{\sqrt{2}} l_{\mu} (x) j^{\mu} (x),
\end{equation}
($\mu = 0, 1, 2, 3$), where $G=G_F \cos \theta_c$ ($G_F \approx 8.975 \times 10^{-5}$ MeV fm$^3$ is the Fermi weak interaction constant; $\cos \theta_c = 0.974$ is the cosine of the Cabibbo angle), $l_{\mu}$ is the lepton weak charged current, and $j^{\mu}$ is the weak current of the nucleons. 
The lepton and nucleon weak charged currents have the forms
\begin{equation}\label{eq:lepton weak neutral current}
l_{\mu} (x) =\hat{\bar{\psi}}_{l^{\prime}}^{} \gamma_{\mu} (1-\gamma_5) \hat{\psi}_{l}^{},
\end{equation}
and
\begin{align}\label{eq:nucleon weak current}
j^{\mu} (x) & = \hat{\bar{\psi}}^{}_{\tauv^{\prime}} \gamma^{\mu} (g^{}_{V} - g^{}_{A} \gamma^{5}) \hat{\psi}^{}_{\tauv},
\end{align}
where $\hat{\psi}_{l}, \hat{\psi}_{l^{\prime}}, \hat{\psi}_{\tauv}, \hat{\psi}_{\tauv^{\prime}}$ represent the corresponding field operators, $\hat{\bar{\psi}} = \hat{\psi}^\dagger \gamma^0$, and $g^{}_{V}=1$ and $g^{}_{A}=1.23$ \citep{Reddy1998} are the relevant charged current vector and axial-vector coupling constants.
$\gamma_{\mu}$ are the usual $\gamma$ matrices and $\gamma^{5} = i \gamma^0 \gamma^{1} \gamma^2 \gamma^3 =\gamma_5$ \citep{BjorkenDrell}.

We can see that the above procedure is similar to the scattering case of Ref. \cite{Duan2023}.
Therefore, we do not repeat the details of the non-relativistic reduction of the field operators here.
Then $j^{\mu}(x)$ can be written as
\begin{align}\label{eq:nucleon weak current-approx}
j^{\mu} (x) \approx (g^{}_{V} \hat{\psi}_{\tauv^{\prime}}^{\dagger} \hat{\psi}^{}_{\tauv}, - g^{}_{A} \hat{\psi}_{\tauv^{\prime}}^{\dagger} \sigmav \hat{\psi}^{}_{\tauv^{}} )^{\mu},
\end{align}
with $\sigmav$ the Pauli matrices.
The interaction Hamiltonian can be given by using 
\begin{equation}\label{eq:interaction hamiltonian at finite}
\hat{H}_{I}  = \int d^3 x \mathcal{H}_{I} = - \int d^3 x \mathcal{L}_{I}.
\end{equation}
Here we take the electron capture process $e^- + p \rightarrow \nu_e + n$ as an example and sandwich the interaction Hamiltonian between the final and initial states to obtain
\begin{equation}\label{eq:Hfi}
H_{fi}= \langle \pv_{\nu}, N+1, Z-1, \lambda^{\prime} \vert \hat{H}_{I}^{} \vert \pv_e, s_e, N, Z, \lambda \rangle,
\end{equation}
where $\pv_e$ and $\pv_{\nu}$ denote the electron and neutrino momenta, $s_e$ is the electron spin, $\{N, Z, \lambda\}$ and $\{ N+1, Z-1, \lambda^{\prime} \}$ characterize the initial and final nuclear-matter states, respectively.
According to Fermi's golden rule \citep{SakuraiNapolitano}, the transition rate is given by
\begin{align}\label{eq:fermi's golden rule}
  R_{\pv_e, \lambda \rightarrow \pv_{\nu}, \lambda^{\prime}}
    & = \frac{1}{2} \sum_{s_e} R_{\pv_e, s_e, \lambda \rightarrow \pv_{\nu}, \lambda^{\prime}} \nonumber \\
  & =\frac{1}{2} \sum_{s_e} 2\pi \vert H_{fi} \vert^2 \delta(E_f-E_i) (1-n_{\pv_{\nu}}^{\nu})\,,
\end{align}
where we averaged over the spin of the incoming electron and the  $1-n_{\pv_{\nu}}^{\nu}$ factor accounts for the Pauli blocking of the final neutrino state.
Using $E_i = E_{\lambda}+E_e$ and $E_f = E_{\lambda'}+E_{\nu}$ ($E_{e}$ and $E_{\nu}$ are the electron and neutrino energies and $E_{\lambda}$ and $E_{\lambda'}$ are the initial and final energies of the nuclear matter), we can write the rate as
\begin{equation}
  R_{\pv_e, \lambda \rightarrow \pv_{\nu}, \lambda^{\prime}}
  =\frac{1}{2} \sum_{s_e} 2\pi \vert H_{fi} \vert^2 \delta( \omega -E_{\lambda^{\prime} \lambda})(1-n_{\pv_{\nu}}^{\nu}),
\end{equation}
where $\omega = E_{e} - E_{\nu}$ is the energy transfer and $E_{\lambda^{\prime} \lambda} = E_{\lambda^{\prime}} - E_{\lambda}$.
Taking a statistical average over initial nuclear-matter states and summing over final nuclear-matter states, we can calculate the neutrino production rate using the following expression:
\begin{widetext}

\begin{multline}\label{eq:rate-initial state to final state}
  R_{\pv_{e} \rightarrow \pv_{\nu}}
  = \sum_{N, Z, \lambda, \lambda^{\prime}} \frac{1}{\mathcal{Z}} e^{-(E_{\lambda}-\mu_n N-\mu_p Z)/T} R_{\pv_{e}, \lambda \rightarrow \pv_{\nu}, \lambda^{\prime}}
  = \frac{G^2 \pi}{V} \Big\{ \Big(1+ \frac{p_{e}}{E_{e}} \cos \theta \Big) g^{2}_{V} S_{np}^{(S=0)} (\qv,\omega) \\
  + \left[\frac{p_e}{E_e} \hat{p}_{\nu i} \hat{p}_{e j}
    + \frac{p_e}{E_e} \hat{p}_{\nu j} \hat{p}_{e i}
    + \Big(1-\frac{p_e}{E_e} \cos\theta \Big) \delta^{}_{ij}
    + i \varepsilon^{}_{ijk} \Big(\frac{p_e}{E_e} \hat{p}_{e k} - \hat{p}_{\nu k} \Big) \right]
  g^{2}_{A} S_{np,ij}^{(S=1)}(\qv,\omega)\Big\} (1-n_{\pv_{\nu}}^{\nu}),
\end{multline}
where $\mathcal{Z}$ is the partition function, $\mu_n$ and $\mu_p$ are neutron and proton chemical potentials, $T$ is the temperature, summation over repeated indices is implied ($i,j,k = 1\dots 3$), $\hat{\pv}_e$ and $\hat{\pv}_{\nu}$ are unit vectors in direction of $\pv_e$ and $\pv_{\nu}$, respectively, $\qv = \pv_{e} - \pv_{\nu}$ is the momentum transfer and $\theta$ is the angle between $\pv_{e}$ and $\pv_{\nu}$.
We have temporarily introduced a finite system volume $V$ but we will take the continuum limit afterwards.
To express the spin-0 $(S=0)$ and spin-1 $(S=1)$ dynamic structure factors in a single equation, we introduce the notation $\sigma_0=1$, $i=j=0$ for $S=0$ and $i,j=1,2,3$ for $S=1$.
Then, the dynamical structure factors are defined as
\begin{multline}
\label{eq:structure factor S-s1}
S_{np,ij} (\qv ,\omega) = \frac{1}{V} \sum_{N, Z, \lambda, \lambda^{\prime}} \frac{1}{\mathcal{Z}} e^{-(E_{\lambda}-\mu_n N-\mu_p Z)/T} \int_{V} d^3 x' e^{-i \qv \cdot \vect{x}'} \int_{V} d^3 x e^{i \qv \cdot \vect{x}} \\
\times \langle N, Z, \lambda \vert   \hat{\psi}^{\dagger}_{p} (\vect{x}') \sigma_{j} \hat{\psi}_{n}(\vect{x}')  \vert  N+1, Z-1, \lambda^{\prime} \rangle \langle N+1, Z-1, \lambda^{\prime} \vert   \hat{\psi}^{\dagger}_{n} (\vect{x}) \sigma_{i} \hat{\psi}_{p}^{}(\vect{x})  \vert N, Z, \lambda \rangle \delta(\omega -E_{\lambda^{\prime} \lambda})\,.
\end{multline}
They can be written in the following form
\begin{equation}\label{eq:structure factors and correlation functions}
S_{np,ij}(\qv, \omega) = - \frac{1}{\pi} \Big[1+ \frac{1}{e^{(\omega-\mu_{n}+\mu_{p})/T}-1}\Big] \Im \Pi_{np,ij}^{R} (\vect{q}, \omega),
\end{equation}
where $\Pi_{np,ij}^{R} (\vect{q}, \omega)$ is the Fourier transform of the retarded correlation function which is defined as
\begin{equation}\label{eq:definition of correlation functions}
  \Pi_{np,ij}^{R} (\vect{x},t)
  = -i \theta(t) \sum_{N, Z, \lambda} \frac{1}{\mathcal{Z}} e^{-(E_{\lambda}-\mu_n N-\mu_p Z)/T}
    \langle N, Z, \lambda |
      [ \hat{\psi}_{p}^{\dagger} (\vect{x},t) \sigma_{j}  \hat{\psi}_{n}^{} (\vect{x},t),
        \hat{\psi}_{n}^{\dagger} (\vect{0},0) \sigma_{i} \hat{\psi}_{p}^{} (\vect{0},0)]
    | N, Z, \lambda \rangle\,.
\end{equation}
Thus we can relate the neutrino production rate to the nuclear density and spin-density correlation functions.

As in the scattering case, the $\varepsilon^{}_{ijk}$ term in Eq. \eqref{eq:rate-initial state to final state} does not contribute.
After the standard projection of $S^{(S=1)}_{np,ij}$ onto transverse ($M=\pm 1$) and longitudinal $(M=0)$ components (cf., e.g., Ref.~\cite{Duan2023} or Eqs. \eqref{eq:Pi-s1-m1} and \eqref{eq:Pi-s1-m0} below), the neutrino production rate can be written as
\begin{equation}\label{eq:neutrino-production-rate-electron}
R_{\produ}  =2 \frac{V^2}{(2\pi)^6} \int d^3 p_{\nu} \int d^3 p_{e} n_{\pv_{e}}^{e} R_{\pv_{e} \rightarrow \pv_{\nu}}.
\end{equation}
So the corresponding differential rate $\frac{dR}{d^3p_\nu d^3 x}$ is
\begin{align}\label{eq:differential-neutrino-production-rate-electron}
  \frac{dR_{\produ}}{d^3p_\nu d^3 x}
  = & \frac{2V}{(2 \pi)^6} \int d^3 p_{e}^{}  n_{\pv_e}^{e} R_{\pv_{e} \rightarrow \pv_{\nu}} \nonumber\\
  = & \frac{G^2}{(2 \pi)^4} \int_{q_{\tmin}}^{q_{\tmax}} dq \int_{\omega_{\tmin}}^{\omega_{\tmax}} d \omega
    \frac{E_e}{E_{\nu}} q n_{\pv_e}^{e}
    \Big\{ \Big(1+ \frac{p_e}{E_e}\cos \theta \Big) g_{V}^2 S_{np}^{(S=0)} (\qv, \omega)
      + \Big(3-\frac{p_e}{E_e} \cos \theta \Big)g_A^2 S_{np}^{(SM=11)} (\qv,\omega) \nonumber \\
    & + \Big(1+\frac{p_e}{E_e} \cos \theta-\frac{2p_e^2 E_\nu}{E_eq^2} \sin^2 \theta \Big) g_A^2
      \Big[S_{np}^{(SM=10)}(\qv,\omega)-S_{np}^{(SM=11)}(\qv,\omega)\Big]
    \Big\} (1-n_{\pv_{\nu}}^{\nu}).
\end{align}
We can see that the volume $V$ drops out after the above procedure.
Notice that the last term (last line) in Eq.~\eqref{eq:differential-neutrino-production-rate-electron} is usually not given in the literature and it is indeed very small.
It stems from the small difference between longitudinal and transverse spin responses due to the spin-orbit coupling.

Equation \eqref{eq:differential-neutrino-production-rate-electron} can be generalized to all processes described by Eqs. \eqref{eq:charged-current processes1} and \eqref{eq:charged-current processes2}, and to cases in which muons instead of electrons interact with nuclear matter.
Namely, the differential rates of (anti)neutrino production or absorption rates can be generically written as
\begin{align}\label{eq:differential rates for all processes}
 \frac{dR}{d^3p_{\nu} d^3 x}  
 = & \frac{G^2}{(2 \pi)^4} \int_{q_{\tmin}}^{q_{\tmax}} dq \int_{\omega_{\tmin}}^{\omega_{\tmax}} d \omega  q
   \frac{E_l}{E_{\nu}} F_l F_{\nu}
   \Big\{ \Big(1+ \frac{p_l}{E_l} \cos \theta \Big) g_{V}^2 S_{\tauv^{\prime} \tauv}^{(S=0)} (\qv, \omega)
     +\Big(3- \frac{p_l}{E_l} \cos \theta \Big)g_A^2 S_{\tauv^{\prime} \tauv}^{(SM=11)} (\qv,\omega) \nonumber \\
   & + \Big(1+\frac{p_l}{E_l} \cos \theta+\Sign_l \Sign_{\nu} \frac{2p_l^2 E_{\nu}}{E_l q^2} \sin^2 \theta \Big)
     g_A^2 \Big[S_{\tauv^{\prime}\tauv}^{(SM=10)}(\qv,\omega)-S_{\tauv^{\prime} \tauv}^{(SM=11)}(\qv,\omega)\Big]
   \Big\},
\end{align}
\end{widetext}
where $l$ denotes one of the charged leptons $e^-$, $e^+$, $\mu^-$, or $\mu^+$, $\nu$ denotes either a neutrino or an antineutrino of the corresponding flavour, and the signs $\Sign_l$ and $\Sign_\nu$ encode whether the corresponding particle $i (= l,\nu)$ is incoming ($\Sign_i = 1$) or outgoing ($\Sign_i=-1$). Hence, momentum and energy transfer are given by $\qv= \Sign_l \pv_{l} +\Sign_{\nu} \pv_{\nu}$, $\omega=\Sign_l E_l + \Sign_{\nu} E_{\nu}$.
For the charged leptons, we use $p_l=\sqrt{E_l^2-m_l^2}$, while the neutrino mass is of course negligible ($p_\nu = E_\nu$).
$F_l$ and $F_\nu$ are generic notations for the occupation numbers $n_{\pv_i}$ for incoming particles $i$ or Pauli blocking factors $(1-n_{\pv_i})$ for outgoing particles $i$. For the charged leptons, we can assume an equilibrium distribution and hence $F_l = 1/(e^{\Sign_l(E_l-\mu_l)/T}+1)$, where the chemical potentials of leptons and antileptons are related to each other by $\mu_{\overline{l}} = -\mu_l$.

\begin{table*}
\centering
\caption{Expressions of the subscripts $\tauv$, $\tauv^{\prime}$, $l$, and $\nu$, and of the factors $F_l$, $F_{\nu}$, $\Sign_l$, and $\Sign_{\nu}$ for the electron (positron) processes.
\label{table:expressions of factors}}
\begin{ruledtabular}
\begin{tabular}{ccccccccc}

processes&$\tauv$ & $\tauv^{\prime}$ & $l$ & $\nu$ & $F_l$ & $F_{\nu}$ & $\Sign_l$ & $\Sign_{\nu}$ \\
\hline
$e^- + p \rightarrow \nu_e + n$ & $p$ & $n$ & $e^-$ & $\nu_e$ & $n_{\pv_e}^{e}$ & $1-n_{\pv_{\nu}}^{\nu}$ & 1 & -1   \\
$\nu_e + n \rightarrow e^- + p $ & $n$ & $p$ & $e^-$ & $\nu_e$ & $1-n_{\pv_e}^{e}$ & $n_{\pv_{\nu}}^{\nu}$ & -1 & 1   \\
$ n \rightarrow  p + e^- + \nubar_e$ & $n$ & $p$ & $e^-$ & $\nubar_e$ & $1-n_{\pv_e}^{e}$ & $1-n_{\pv_{\nu}}^{\nu}$ & -1 & -1   \\
$p + e^- + \nubar_e \rightarrow n$ & $p$ & $n$ & $e^-$ & $\nubar_e$ & $n_{\pv_e}^{e}$ & $n_{\pv_{\nu}}^{\nu}$ & 1 & 1   \\
$p \rightarrow n + e^+ + \nu_e$ & $p$ & $n$ & $e^+$ & $\nu_e$ & $1-n_{\pv_{e}}^{e}$ & $1-n_{\pv_{\nu}}^{\nu}$ & -1 & -1   \\
$n + e^+ + \nu_e \rightarrow p $ & $n$ & $p$ & $e^+$ & $\nu_e$ & $n_{\pv_{e}}^{e}$ & $n_{\pv_{\nu}}^{\nu}$ & 1 & 1   \\
$e^+ + n \rightarrow \nubar_e + p$ & $n$ & $p$ & $e^+$ & $\nubar_e$ & $n_{\pv_{e}}^{e}$ & $1-n_{\pv_{\nu}}^{\nu}$ & 1 & -1   \\
$\nubar_e + p \rightarrow e^+ + n$ & $p$ & $n$ & $e^+$ & $\nubar_e$ & $1-n_{\pv_{e}}^{e}$ & $n_{\pv_{\nu}}^{\nu}$ & -1 & 1  \\

\end{tabular}
\end{ruledtabular}
\end{table*}
To compute the integrals, $E_l$ and $\cos \theta$ should be expressed in terms of $q$ and $\omega$, i.e.,
\begin{align}\label{eq:E and costheta}
  E_l &=\Sign_l(\omega - \Sign_{\nu} E_{\nu}),\\
  \cos \theta &=\Sign_l \Sign_{\nu} \frac{q^2-p_l^2-E_{\nu}^2}{2 p_l E_{\nu}}.
\end{align}
Besides, the integration limits have expressions as follows:
\begin{align}\label{eq:integration-limits-omega}
& \omega_{\tmin} = \Sign_l \sqrt{(q-\Sign_lE_{\nu})^2 + m_l^2} + \Sign_{\nu} E_{\nu}, \nonumber \\
& \omega_{\tmax} = \Sign_l \sqrt{(q+\Sign_lE_{\nu})^2 + m_l^2} + \Sign_{\nu} E_{\nu}.
\end{align}
In the cases with an incoming charged lepton, we can in practice neglect the lepton occupation numbers beyond some maximum lepton momentum $p_{l,\tmax}$ and hence
\begin{align}\label{eq:integration-limits-q-1}
& q_{\tmin} =0, \nonumber \\
& q_{\tmax}=p_{l,\tmax} + p_{\nu}.
\end{align}
In the cases with an outgoing charged lepton, Pauli blocking may, at low temperatures, require the outgoing lepton momentum to lie above some minimum value $p_{l,\tmin}$, and then 
\begin{align}\label{eq:integration-limits-q-2}
& q_{\tmin} =\max \{0,p_{l,\tmin}-p_{\nu} \}, \nonumber \\
& q_{\tmax}=\infty.
\end{align}
As an example, expressions of the subscripts $\tauv$, $\tauv^{\prime}$, $l$, and $\nu$, and of the factors $F_l$, $F_{\nu}$, $\Sign_l$, and $\Sign_{\nu}$ in the cases in which the lepton is an electron or a positron, are listed in Table \ref{table:expressions of factors}.
Analogous expressions can be obtained in the muon cases after replacing $e$ with $\mu$.

For the structure factor $S_{pn}$ needed in some of these processes, one has to replace in Eq. \eqref{eq:structure factor S-s1} $n\leftrightarrow p$, $N+1\to N-1$, $Z-1\to Z+1$. However, after relabeling in the resulting equation the summation variables $N\to N+1$, $Z\to Z-1$, it is straight-forward to show that
\begin{align}\label{eq:Spn-Snp}
S_{pn} (q,\omega) = e^{(\omega+\mu_n-\mu_p)/T} S_{np} (q,-\omega).
\end{align}
This means that once one of the two structure factors is computed, the other one can easily be obtained.

\subsection{Charged-current RPA response functions with Skyrme interaction in asymmetric nuclear matter} 
\label{subsec:response functions in ANM}
Skyrme interactions are well-known and widely used effective nucleon-nucleon interactions to solve the nuclear many-body problem.
Initially, they were mostly employed in nuclear structure calculations \citep{Vautherin1972,Beiner1975,Kohler1976}.
The corresponding functionals are also widely used to describe nuclear matter, such as neutron star matter.
With the notations of Ref. \citep{Bender2003}, the isovector part of the Skyrme energy-density functional can be written as
\begin{align}
\varepsilon^{}_{\text{Sk,IV}} = 
& C_1^{\rho} \vec{\rho}_1^{\ 2} + C_1^{\tau} (\vec{\rho}^{}_1\! \cdot\! \vec{\tau}^{}_1 - \vec{\vect{j}}{}_1^{\ 2}) 
+ C_1^{\laplacian \rho} \vec{\rho}^{}_1\! \cdot\! \laplacian \vec{\rho}^{}_1 \nonumber \\
& + C_1^{\nabla\rho} (\nablav \vec{\rho}^{}_1)^2 
+ C_1^{s} \vec{\sv}_1^2  + C_1^{sT} ( \vec{\sv}^{}_1\! \cdot\! \vec{\vect{T}}^{}_1 - \vec{\mathbbm{J}}_1^2) \nonumber\\
& + C_1^{\laplacian s} \vec{\sv}^{}_1\! \cdot\! \laplacian \vec{\sv}^{}_1  
+ C_1^{\nabla\otimes s} (\nablav\!\otimes\! \vec{\sv}^{}_1)^2 
+ C_1^{\nabla s} (\nablav\!\cdot\! \vec{\sv}^{}_1)^2 \nonumber \\
& + C_1^{\nabla J} [\vec{\rho}^{}_1\! \cdot\! (\nablav\! \cdot\! \vec{\vect{J}}^{}_1) + \vec{\sv}^{}_1\! \cdot\! \nablav \!\times\! \vec{\vect{j}}^{}_1]\,. 
\end{align}
Here we have included additional $(\nablav \rho)^2$ and $(\nablav\otimes\sv)^2 \equiv \sum_{ij} (\nabla_i s_j)^2$ terms (cf. Ref. \cite{Duan2023}) that are absent in Ref. \cite{Bender2003} but that are needed in the case of generalized Skyrme functionals.
Notice that, in the charged-current case, we define
\begin{equation}\label{eq:tau-plus-minus}
  \tau^{}_{\pm}= \frac{1}{2} (\tau^{}_1 \pm i \tau^{}_2),
\end{equation}
where $\tau^{}_i$ ($i=1,2,3$) are the isospin Pauli matrices.
Then, besides the $\tau^{}_3$ contribution to $\vec{\rho}_1^{\ 2}$ as in Eq. (20) of Ref. \cite{Duan2024}, there are also $\tau^{}_{+}$ and $\tau^{}_{-}$ contributions, i.e., $\langle \hat{\psi}^{\dagger} \tau^{}_{+} \hat{\psi} \rangle \langle \hat{\psi}^{\dagger} \tau^{}_{-} \hat{\psi} \rangle = \langle \hat{\psi}_n^{\dagger} \hat{\psi}_p \rangle \langle \hat{\psi}_p^{\dagger} \hat{\psi}_n \rangle$, and analogously for other terms.
Thus, there are charge exchanges.
Then the energy functional is given by integration
\begin{equation}\label{eq:energy functional}
E_{\text{Sk,IV}} = \int d^3 r \varepsilon^{}_{\text{Sk,IV}}.
\end{equation}

In this work, we will compute the response $\Pi_{np}$ needed in Eq. \eqref{eq:structure factors and correlation functions} using Skyrme functionals within the RPA.
Expressions for full Skyrme RPA response functions in the charge-exchange channels were given in Ref. \cite{Davesne2019}. Here we will use an alternative but equivalent approach.
Following the notation in Ref. \citep{Urban2020}, the residual particle-hole (ph) interaction is derived by computing 
\begin{equation}\label{eq:ph-interaction}
  \mathcal{V}_{21}^0 = \frac{\delta ^2 E_{\rm{Sk,IV}}}{\delta \rho^{}_{2' 2} \delta \rho^{}_{1 1'}},
\end{equation}
with the short-hand notation
\begin{align}\label{eq:notation1}
1 =\Big(p, \pv^{}_1+\frac{\qv}{2},s^{}_1\Big),&& 1' =\Big(n, \pv^{}_1-\frac{\qv}{2},s'_1\Big), 
\end{align}
and analogous for $2$ and $2'$. $\rho^{}_{1 1'} = \langle c_{1'}^{\dagger} c^{}_1 \rangle$ denotes the density matrix (which is now non-diagonal in isospin indices).
As in the neutral-current case in asymmetric nuclear matter \citep{Duan2023}, the ph interaction still has the form
\begin{align}\label{eq:ph interaction-ANM}
\mathcal{V}_{21}^0 = & v_1^0 (q) + v_2^0 (p_1^2 +  p_2^2) + v_3^0 \pv_1 \cdot \pv_2 \nonumber \\
& +[ v_4^0 (q) + v_5^0 (p_1^2 + p_2^2 ) + v_6^0 \pv_1 \cdot \pv_2 ] \sigmav_1 \cdot \sigmav_2 \nonumber \\
& + v_8^0 i \qv \cdot (\pv_1 - \pv_2) \times (\sigmav_1 + \sigmav_2),
\end{align}
where the coefficients $v_i^0$ are now given by
\begin{align}\label{eq:vi0}
& v_1^0(q)=4C_1^{\rho}+(-C_1^{\tau}-4C_1^{\Delta \rho}+4C_1^{\nabla \rho}) q^2, \nonumber \\
& v_2^0=2C_1^{\tau}, \nonumber \\
& v_3^0=-4C_1^{\tau}, \nonumber \\
& v_4^0(q)=4C_1^s+(-C_1^{sT}-4C_1^{\Delta s}+4C_1^{\nabla \otimes s})q^2, \nonumber \\
& v_5^0=2C_1^{sT}, \nonumber \\
& v_6^0=-4C_1^{sT}, \nonumber \\
& v_8^0=2C_1^{\nabla J}.
\end{align}
We can see that the expressions of $v_i^0$ here are simpler than in the neutral-current case.
However, as we will see below, the complete procedure is more complicated than in the neutral-current case because of the generalized Lindhard functions.
To obtain the RPA vertex $\mathcal{V}$, we need to solve the Bethe-Salpeter-like equation:
\begin{equation}\label{eq:BS-like equation}
\mathcal{V}_{21} = \mathcal{V}_{21}^0 - \sum_3 \mathcal{V}_{23}^0 G_{\ph} (\pv_3, \qv) \mathcal{V}_{31},
\end{equation}
where $\sum_3 = \sum_{s^{}_3s'_3} \int d^3 p^{}_3/(2\pi)^3$, and the neutron-particle proton-hole Green's function [corresponding to the $np$ channel in Eq. \eqref{eq:structure factor S-s1}] is defined as
\begin{equation}\label{eq:ph green's function}
G_{\ph} (\pv,\qv, \omega) = \frac{n_{ \pv + \frac{\qv}{2}}^{n} - n_{ \pv -\frac{\qv}{2}}^{p}}{\omega - (  \epsilon_{ \pv +\frac{\qv}{2}}^{n} - \epsilon_{ \pv -\frac{\qv}{2}}^{p}  ) +i \eta},
\end{equation}
where $n^{\tauv}_{\pv} =
1/(e^{(\epsilon^{\tauv}_{\pv}-\mu_{\tauv})/T}+1)$ denotes the
finite-temperature occupation number and $\epsilon^{\tauv}_{\pv} =
\frac{p^2}{2m^*_{\tauv}}+U_{\tauv}$ the Hartree-Fock (HF)
single-particle energy, with $\mu_{\tauv}$ the chemical potential,
$U_{\tauv}$ the mean field, and $m^*_{\tauv}$ the effective mass of
nucleons of kind $\tauv$.

As in Ref. \cite{Duan2023}, the RPA vertex is finally determined as
\begin{align}\label{eq:RPA vertex-ANM}
\mathcal{V}_{21} = & v_1 +v_2 p_1^2  + v_3 \pv_1 \ncdot \pv_2 + v_4 \sigmav_1 \ncdot \sigmav_2  + v_5 \sigmav_1 \ncdot \sigmav_2 p_1^2 \nonumber\\
& + v_6 \sigmav_1 \ncdot \sigmav_2 \pv_1 \ncdot \pv_2  +v_7 \sigmav_1 \ncdot \qv \sigmav_2 \ncdot \qv  + v_8 i \qv \ncdot \pv_1 \ntimes \sigmav_1  \nonumber\\
& + v_9 i \qv \ncdot \pv_1 \ntimes \sigmav_2  + v_{10} p_1^2 p_2^2  + v_{11} \pv_1 \ncdot \qv \pv_2 \ncdot \qv \nonumber\\
& + v_{12} \sigmav_1 \ncdot \sigmav_2 p_1^2 p_2^2  + v_{13} \sigmav_1 \ncdot \sigmav_2 \pv_1 \ncdot \qv \pv_2 \ncdot \qv \nonumber\\
& + v_{14} \sigmav_1 \ncdot \qv \sigmav_2 \ncdot \qv p_1^2  +v_{15} \sigmav_1 \ncdot \qv \sigmav_2 \ncdot \qv p_1^2 p_2^2  \nonumber\\
& - v_{16} i \qv \ncdot  \pv_2 \ntimes \sigmav_2 p_1^2  - v_{17} i \qv \ncdot  \pv_2 \ntimes \sigmav_1 p_1^2 \nonumber\\
& + v_{18} \qv \ncdot \pv_1 \ntimes \sigmav_1 \qv \ncdot \pv_2 \ntimes \sigmav_2 + v_{19} \pv_1 \ncdot \qv \nonumber\\
& + v_{20} \pv_1 \ncdot \qv p_2^2  + v_{21} \sigmav_1 \ncdot \sigmav_2  \pv_1 \ncdot \qv  + v_{22}  \sigmav_1 \ncdot \sigmav_2  \pv_1 \ncdot \qv p_2^2  \nonumber\\
& + v_{23} \sigmav_1 \ncdot \qv \sigmav_2 \ncdot \qv \pv_1 \ncdot \qv  + v_{24}  \sigmav_1 \ncdot \qv \sigmav_2 \ncdot \qv \pv_1 \ncdot \qv p_2^2 \nonumber\\
& + v_{25} \sigmav_1 \ncdot \qv \sigmav_2 \ncdot \qv \pv_1 \ncdot \qv \pv_2 \ncdot \qv  - v_{26} i \qv \ncdot \pv_2 \ntimes \sigmav_2  \pv_1 \ncdot \qv \nonumber\\
& - v_{27} i \qv \ncdot  \pv_2 \ntimes \sigmav_1 \pv_1 \ncdot \qv  +v_{28} p_2^2  + v_{29} \sigmav_1 \ncdot \sigmav_2 p_2^2 \nonumber\\
& -v_{30} i \qv \ncdot \pv_2 \ntimes \sigmav_2 - v_{31} i \qv \ncdot \pv_2 \ntimes \sigmav_1 + v_{32} \sigmav_1 \ncdot \qv \sigmav_2 \ncdot \qv p_2^2 \nonumber\\
& + v_{33} i \qv \ncdot  \pv_1 \ntimes \sigmav_1 p_2^2 +  v_{34} i \qv \ncdot  \pv_1 \ntimes \sigmav_2 p_2^2  + v_{35} \pv_2 \ncdot \qv \nonumber\\
& + v_{36} \pv_2 \ncdot \qv p_1^2 + v_{37} \sigmav_1 \ncdot \sigmav_2  \pv_2 \ncdot \qv + v_{38}  \sigmav_1 \ncdot \sigmav_2  \pv_2 \ncdot \qv p_1^2 \nonumber\\
& + v_{39} \sigmav_1 \ncdot \qv \sigmav_2 \ncdot \qv \pv_2 \ncdot \qv + v_{40}  \sigmav_1 \ncdot \qv \sigmav_2 \ncdot \qv \pv_2 \ncdot \qv p_1^2 \nonumber\\
& +v_{41} i \qv \ncdot \pv_1 \ntimes \sigmav_1  \pv_2 \ncdot \qv + v_{42} i \qv \ncdot  \pv_1 \ntimes \sigmav_2 \pv_2 \ncdot \qv.
\end{align}
However, to compute the coefficients $v_i$ following the steps explained in Ref. \cite{Urban2020}, it is not enough anymore to define only the three generalized Lindhard functions
\begin{align}\label{eq:lindhard functions}
 \Pi_{k}(q, \omega) &= - 2 \int \frac{d^3 p}{(2\pi)^3} p^{k} G_{\ph} (\pv, \qv, \omega), \\
 \Pi_{2L} (q, \omega) &= - 2 \int \frac{d^3 p}{(2 \pi)^3} p^2 \cos^2 \vartheta \, G_{\ph} (\pv, \qv, \omega), \\
 \Pi_{2T} (q, \omega)  &= \frac{\Pi_2 - \Pi_{2L}}{2},
\end{align}
where $\vartheta$ is the angle between $\pv$ and $\qv$, because Eq. (31) in Ref. \cite{Duan2023} is not valid in the present charged-current case.
Therefore, to complete the procedure, we define two additional functions
\begin{align}\label{eq:PiA}
\Pi_A(q,\omega) &= - 2 \int \frac{d^3 p}{(2\pi)^3} \pv \cdot \qv G_{\ph} (\pv, \qv, \omega),\\
\label{eq:PiB}
\Pi_B(q,\omega) &= - 2 \int \frac{d^3 p}{(2\pi)^3} p^2 \pv \cdot \qv G_{\ph} (\pv, \qv, \omega).
\end{align}
The above generalized Lindhard functions can be written using the $\beta_i$ functions of Refs. \cite{Hernandez1999,Pastore2015}.
Similarly to the neutral-current case in Ref. \cite{Duan2023}, we solve the following linear system of equations
\begin{equation}
  \label{eq:linear-system}
 \sum_k (\delta_{ik}-A_{ik})v_{k} = v_{i}^{0},
\end{equation}
to obtain $v_i$.
Notice that in addition to the $v_i^0$ defined in Eq. \eqref{eq:ph interaction-ANM}, we have $v_{28}^0 = v_2^0, v_{29}^0 = v_5^0, v_9^0=v_{30}^0=v_{31}^0=v_8^0$, and the remaining $v_i^0$ vanish. The expressions for $A_{ik}$ are given in the appendix \ref{sec:expression of Aik}.
They are simpler than in the neutral-current case because here we have only one combination of isospin labels ($np$), while in the neutral-current case there is a coupling between $nn$ and $pp$ channels.

The full RPA response functions can be given after finishing the numerical computation of the coefficients $v_i$ by using the relation
\begin{multline}\label{eq:Pi-s1}
\Pi_{\RPA,np,ij}= -\sum_1 \sigma_{1i} G_{\ph} (\pv_1,\qv) \sigma_{1j} \\
+ \sum_{1,2} \sigma_{2j} G_{\ph}(\pv_2,\qv) \mathcal{V}_{21} G_{\ph} (\pv_1,\qv) \sigma_{1i},
\end{multline}
where $i=j=0$ for $S=0$ and $i,j = 1,2,3$ for $S=1$ as in Eq. \eqref{eq:structure factors and correlation functions}.
After performing the summations in Eq. \eqref{eq:Pi-s1}, one gets the following equations for the density response function ($S=0$),
\begin{align} \label{eq:Pi-s0}
\Pi_{\RPA,np}^{(S=0)}  =  & \Pi_0 + v_{1} \Pi_0^2 + v_{2} \Pi_2 \Pi_0 + \frac{1}{q^2} v_{3} \Pi_A^{2}
+ v_{10} \Pi_2^{2}
\nonumber\\
& +   v_{11} \Pi_A^{2} + v_{19} \Pi_A \Pi_0 +  v_{20} \Pi_A \Pi_2 \nonumber \\
&+ v_{28} \Pi_0 \Pi_2  +  v_{35} \Pi_0 \Pi_A + v_{36} \Pi_2 \Pi_A\,,
\end{align}
the transverse spin response function ($S=1,M=\pm1$),
\begin{align} \label{eq:Pi-s1-m1}
\Pi_{\RPA,np}^{(SM=11)} 
   = &\frac{1}{2} \sum_{i,j=1}^3 \Pi_{\RPA, np, ij}^{(S=1)} \Big(\delta_{ij} - \frac{q_i q_j}{q^2}\Big) 
\nonumber\\
= & \Pi_0 + v_{4} \Pi_0^{2}  + v_{5} \Pi_2 \Pi_0 + \frac{1}{q^2} v_{6} \Pi_A^{2} + v_{12} \Pi_2^{2} \nonumber\\ &
+ v_{13} \Pi_A^{2} +  v_{21} \Pi_A \Pi_0  +  v_{22} \Pi_A \Pi_2 \nonumber \\ &
+ v_{29} \Pi_0 \Pi_2 \!+\!  v_{37} \Pi_0 \Pi_A \! +\! v_{38} \Pi_2 \Pi_A\,,
\end{align}
and the longitudinal spin response function ($S=1,M=0$),
\begin{align} \label{eq:Pi-s1-m0}
\Pi_{\RPA,np}^{(SM=10)}
= & \sum_{i,j=1}^3 \Pi_{\RPA, np,ij}^{(S=1)} \frac{q_i q_j}{q^2}
\nonumber\\
= & \Pi_{\RPA,np}^{(SM=11)} 
+ q^2 \Big(v_{7} \Pi_0^{2} + v_{14} \Pi_2 \Pi_0
+ v_{15} \Pi_2^{2} \nonumber\\
&+  v_{23} \Pi_A \Pi_0 +  v_{24} \Pi_A \Pi_2 + v_{25}\Pi_A^{2} \nonumber\\
& + v_{32} \Pi_0 \Pi_2 +  v_{39} \Pi_0 \Pi_A + v_{40} \Pi_2 \Pi_A \Big).
\end{align}

The lowest-order Landau approximation can be computed when we set $q=0$ and replace $p_1^2$ and $p_2^2$ with $k_{F,n}^2$ and $k_{F,p}^2$ in computing $\mathcal{V}_{21}^0$, where $k_{F,b}=(3\pi^2 \rho_b)^{1/3}$ ($b=n,p$).
The response functions in Landau approximation have the following expressions
\begin{align}\label{eq:Landau-S0}
\Pi_{\Landau, np}^{(S=0)} =\frac{\Pi_0}{1- f_0 \Pi_0},
\end{align}
and
\begin{align}\label{eq:Landau-S1}
\Pi_{\Landau, np}^{(S=1)} =\frac{\Pi_0}{1- g_0 \Pi_0},
\end{align}
with 
\begin{align}\label{eq:spin0}
f_0=v_1^0(q=0) + v_2^0 (k_{F,n}^2 + k_{F,p}^2),
\end{align}

\begin{align}\label{eq:spin1}
g_0=v_4^0(q=0) + v_5^0 (k_{F,n}^2 + k_{F,p}^2).
\end{align}

Once the above $\Pi_{\RPA, np}^{(S,M)}$ or $\Pi_{\Landau, np}^{(S)}$ are obtained, the corresponding dynamical structure factors $S_{np}$ and $S_{pn}$ will be computed from Eqs. \eqref{eq:structure factors and correlation functions} and \eqref{eq:Spn-Snp}.

\section{Results}
\label{sec:results}
As mentioned in the introduction, to solve various problems of previous Skyrme functionals, we recently constructed two alternative Skyrme interactions, Sky3s and Sky4s \citep{Duan2024,Duan2025}, whose effective masses and spin-dependent terms were constrained from microscopic BHF calculations.
Therefore, these interactions may give a better description of (anti)neutrino production and absorption needed for astrophysical simulations.

In Sec.\ref{subsec:results-protoneutron-star} we will study neutrino production and absorption in proto-neutron star and supernova matter.
We did the calculations with both Sky3s and Sky4s, but the results were very similar under the chosen conditions since Sky3s and Sky4s differ mostly by their equation of state at $\rho\gtrsim 0.4$ fm$^{-3}$ (cf. Fig. 4 of Ref. \cite{Duan2024}). Therefore, we present only the results computed with Sky3s. To compare, we also show the results for SLy4. In Sec. \ref{subsec:dUrca}, we will study direct Urca neutrino emission from matter in $\beta$ equilibrium, using exclusively the Sky3s interaction since it has a lower direct Urca threshold than Sky4s and SLy4.

We mention that we tested our code for the calculation of the response
functions by reproducing results obtained with the SLy230a interaction
in Landau approximation in Ref. \cite{Margueron2001}, and with the LNS
interaction in full RPA in Ref. \cite{Dzhioev2018}. Furthermore, we
checked the calculation of the neutrino rates by comparing our SLy4
results in HF and Landau approximation with those of
Ref. \cite{Pascal2022} (HF results correspond to the mean field
approximation in \cite{Pascal2022}).
\subsection{Neutrino production and absorption in proto-neutron star and supernova matter}
\label{subsec:results-protoneutron-star}

\begin{table*}
\centering
\caption{Effective masses, $\Delta U = U_n-U_p$, $\Delta \mu = \mu_n-\mu_p$, chemical potentials of electrons (and muons, when they were taken into account) under the different conditions considered in this work.} 
\label{table:effective-masses-central-potential}
\begin{ruledtabular}
\begin{tabular}{lccccccccccc}
 interaction & $T$  & $\rho$ & $Y_p$ & $Y_e$ & $\beta$ equilibrium & $m^*_n$ & $m^*_p$ & $\Delta U$ & $\Delta \mu$ & $\mu_e$ & $\mu_{\mu}$   \\
 & (MeV) & (fm$^{-3}$) &  &  &  &  (MeV) &  (MeV) &  (MeV) & (MeV) & (MeV) &  (MeV)  \\
 \hline
Sky3s & 19.5  & 0.00498 & 0.0764 & 0.0764 & no & 931.7 & 922.3 & 4.839 & 53.8 & 20.7 &  \\ 
SLy4 & 19.5  & 0.00498 & 0.0764 & 0.0764 & no & 921.7 & 930.9 & 8.177 & 57.8 & 20.7 &  \\ 
Sky3s & 27.6  & 0.256  & 0.069 & 0.069 & no & 781.5 & 655.8 & 67.554 & 169.8 & 143.3 &  \\ 
SLy4 & 27.6   & 0.256  & 0.069 & 0.069 & no & 477.1 & 650.7 & 13.428 & 177.2 & 143.3 &  \\ 
Sky3s & 5 & 0.52 & 0.1655 & 0.0978 &  standard & 727.0 & 635.9 & 135.848 & 225.8 & 225.8 & 225.8 \\ 
Sky3s (RPA) & 5  & 0.52 & 0.1615 & 0.0990 & modified & 727.7 & 635.4 & 137.466 & 228.7 & 226.7 & 221.2 \\
Sky3s (HF) & 5 & 0.52 & 0.1610 & 0.0985 & modified & 727.7 & 635.3 & 137.658 & 229.0 & 226.3 & 221.2 \\

\end{tabular}
\end{ruledtabular}
\end{table*}

\begin{figure*}
\begin{center}
\includegraphics[scale=0.6]{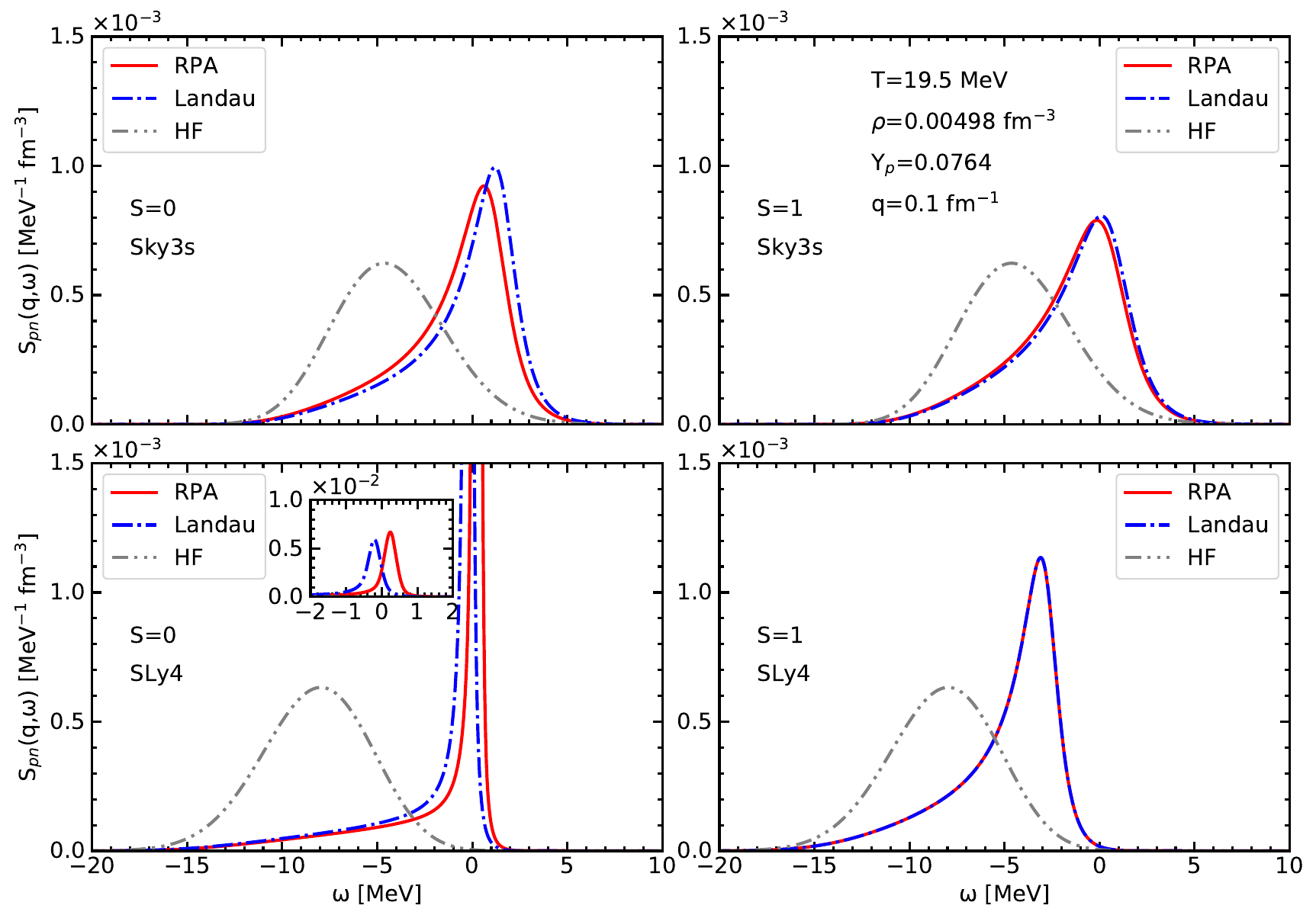} 
\caption{Response functions in HF, Landau approximation, and full RPA for Sky3s (top panels) at $T=19.5$ MeV, $\rho=0.00498$ fm$^{-3}$, $Y_p=0.0764$.
  The momentum transfer is $q=0.1$ fm$^{-1}$. The results for SLy4 (bottom panels) are also shown for comparison.
  Left: $S=0$ responses; right: $S=1$ responses.}
\label{fig:responses-n-p-T19.5}
\end{center}
\end{figure*}

\begin{figure*}
\begin{center}
\includegraphics[scale=0.6]{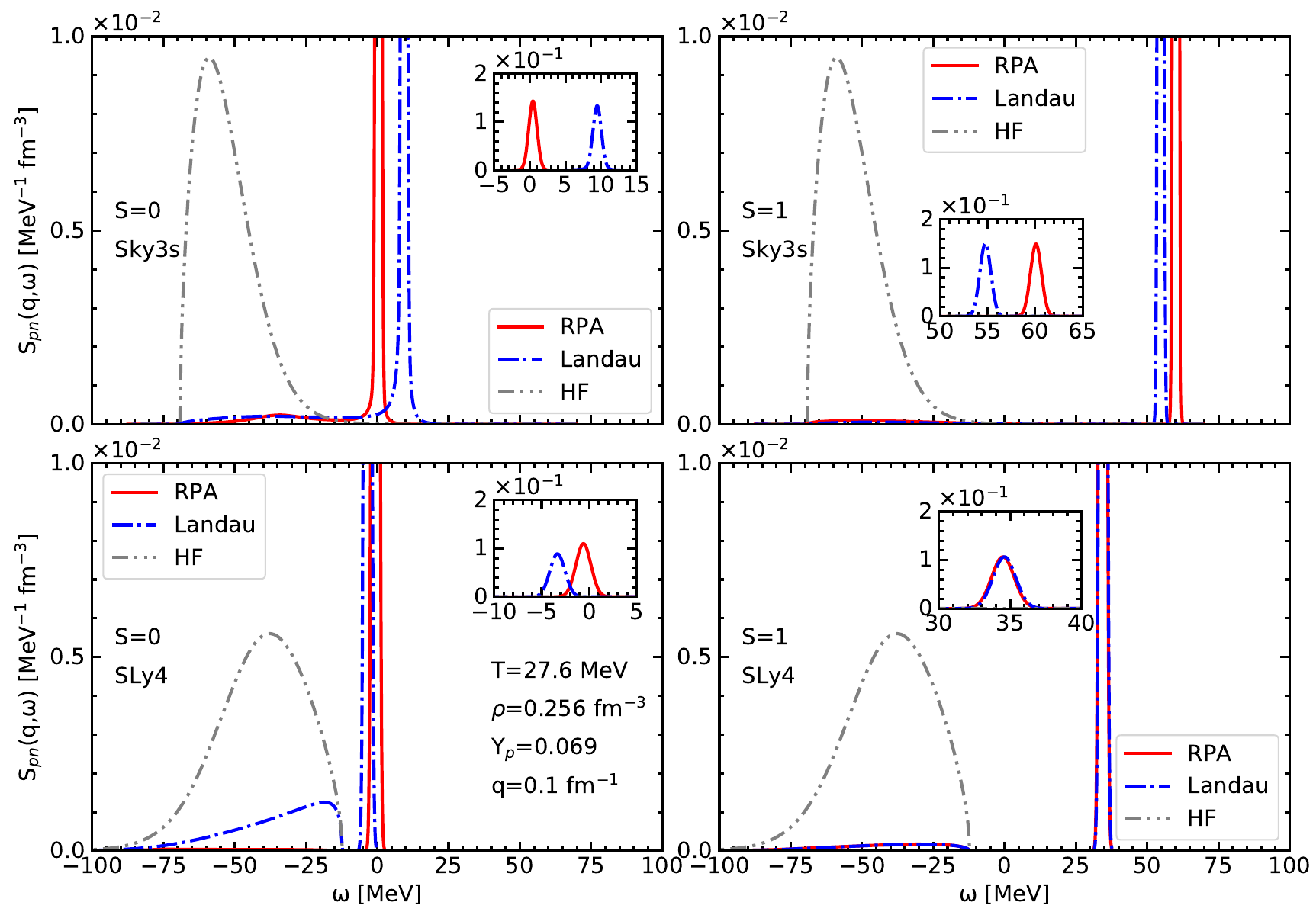} 
\caption{Similar to Fig.~\ref{fig:responses-n-p-T19.5} but for $T=27.6$ MeV, $\rho=0.256$ fm$^{-3}$, $Y_p=0.069$.
}
\label{fig:responses-n-p-T27.6}
\end{center}
\end{figure*}

Like neutrino-nucleon scattering, (anti)neutrino production and absorption are also essential parts of the relevant astrophysical simulations, and they can be related to the charged-current response functions.
Apart from the most important neutrino processes $e^{-} + p \leftrightarrow \nu_e + n$ and $e^{+} + n \leftrightarrow \bar{\nu}_e + p$, also processes $p \leftrightarrow n +e^{+} +\nu_e$ and $n \leftrightarrow p + e^{-} +  \bar{\nu}_e$ were included in simulations in the recent study \cite{Oertel2020,Pascal2022}, where the (anti)neutrino production and absorption were computed at the level of the Landau approximation using the Skyrme functional SLy4 \cite{Chabanat1998}.
Some combinations of temperature $T$, baryon number density $\rho$ (in the astrophysical literature usually called $n_B$), and proton fraction $Y_p$ appearing during the evolution of the proto-neutron star can be found in Table 1 of Ref. \cite{Oertel2020} and Table 1 of Ref. \cite{Pascal2022}.
To compare conveniently, we choose two combinations of $T,\rho,Y_p$ from Refs. \cite{Oertel2020,Pascal2022} for our calculations, although, if one repeated the simulation with Sky3s instead of SLy4, one would probably obtain different density and proton fraction at a given temperature.

As a first example, we choose temperature $T=19.5$ MeV and the corresponding density and proton fraction from Ref. \cite{Oertel2020}.
We can see from Table 1 of Ref. \cite{Oertel2020} that at this low density, matter contains not only neutrons and protons but also some fraction of clusters.
Therefore, we recompute the density and proton fraction of unbound nucleons needed for the computation of the response functions.
Using the parameters given in Ref. \cite{Oertel2020}, we get the density and proton fraction listed in Table \ref{table:effective-masses-central-potential}.
Another combination is from Ref. \cite{Pascal2022}, with temperature $T=27.6$ MeV at a much higher density listed also in Table \ref{table:effective-masses-central-potential}.
These two combinations represent low- and relatively high-density cases.
We do not choose a higher density to avoid that the nucleon Fermi velocity exceeds the speed of light for SLy4 \cite{Duan2023}.
Table \ref{table:effective-masses-central-potential} summarizes also the nucleon effective masses and other quantities of interest such as the difference of mean field potentials $\Delta U = U_n-U_p$, difference of chemical potentials $\Delta \mu = \mu_n-\mu_p$, and chemical potentials of electrons under the different conditions.

First, we present the results of the charged-current response functions needed for the computation of (anti)neutrino production and absorption at the chosen ($T, \rho, Y_p$).
Since $S_{np}$ and $S_{pn}$ are related to each other via Eq. \eqref{eq:Spn-Snp}, we show only $S_{pn} (q,\omega)$ in this work.

Figure \ref{fig:responses-n-p-T19.5} shows the response functions $S_{pn} (q,\omega)$ computed with Sky3s (top panels) at $T=19.5$ MeV, $\rho=0.00498$ fm$^{-3}$, $Y_p=0.0764$, and momentum transfer $q=0.1$ fm$^{-1}$.
In the bottom panels of Fig. \ref{fig:responses-n-p-T19.5}, the results for SLy4 are shown for comparison.
The full RPA, the Landau approximation, and HF responses (i.e., $\Pi_0$) are shown for channels $S=0$ (left panels) and $S=1$ (right panels).
Obviously, the energy transfers $\omega$ of the responses for Sky3s differ from those for SLy4 because these two interactions give different $\Delta U = U_n-U_p$ as can be seen in Table \ref{table:effective-masses-central-potential}.
The strengths of the full RPA and Landau approximation are suppressed at lower energy transfers compared to the HF response, and the missing strength is shifted to higher energy transfers.
In particular, in the $S=0$ channel in SLy4, a large fraction of this strength is concentrated in a narrow peak corresponding to a zero-sound mode, shown in the inset of the lower left panel. In the figure, the peaks were slightly broadened for a better visibility.
Unlike SLy4, Sky3s does not have a zero-sound mode at this momentum transfer.
We can see that there are some differences between the full RPA and Landau approximation, especially in the $S=0$ channel, for both SLy4 and Sky3s. 
Concerning the full RPA for $S=1$, we show only the channel $M=\pm 1$ because the differences between the channels $M=\pm 1$ and $M=0$ are too small under the chosen condition.

\begin{figure*}[ht!]
\includegraphics[scale=0.6]{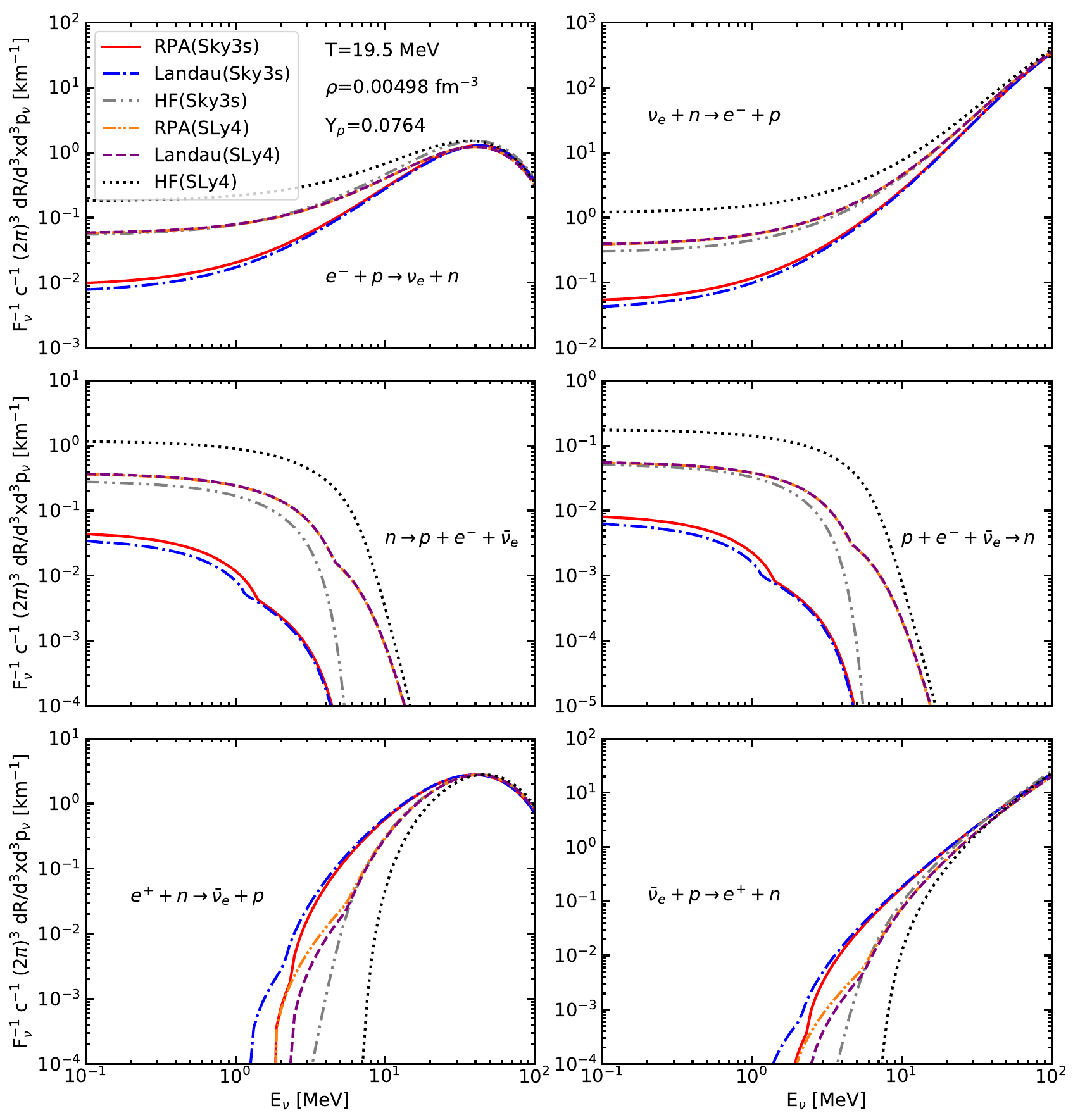} 
\caption{Differential rates of (anti)neutrino production and absorption as a function of the (anti)neutrino energy $E_{\nu}$ for interaction Sky3s at $T=19.5$ MeV, $\rho=0.00498$ fm$^{-3}$, $Y_p=0.0764$.
  The results for SLy4 are also shown for comparison.
  The left panels are for (anti)neutrino production, and the right panels are for (anti)neutrino absorption.}
\label{fig:neutrino-production-absorption-T19.5}
\end{figure*}

\begin{figure*}
\includegraphics[scale=0.6]{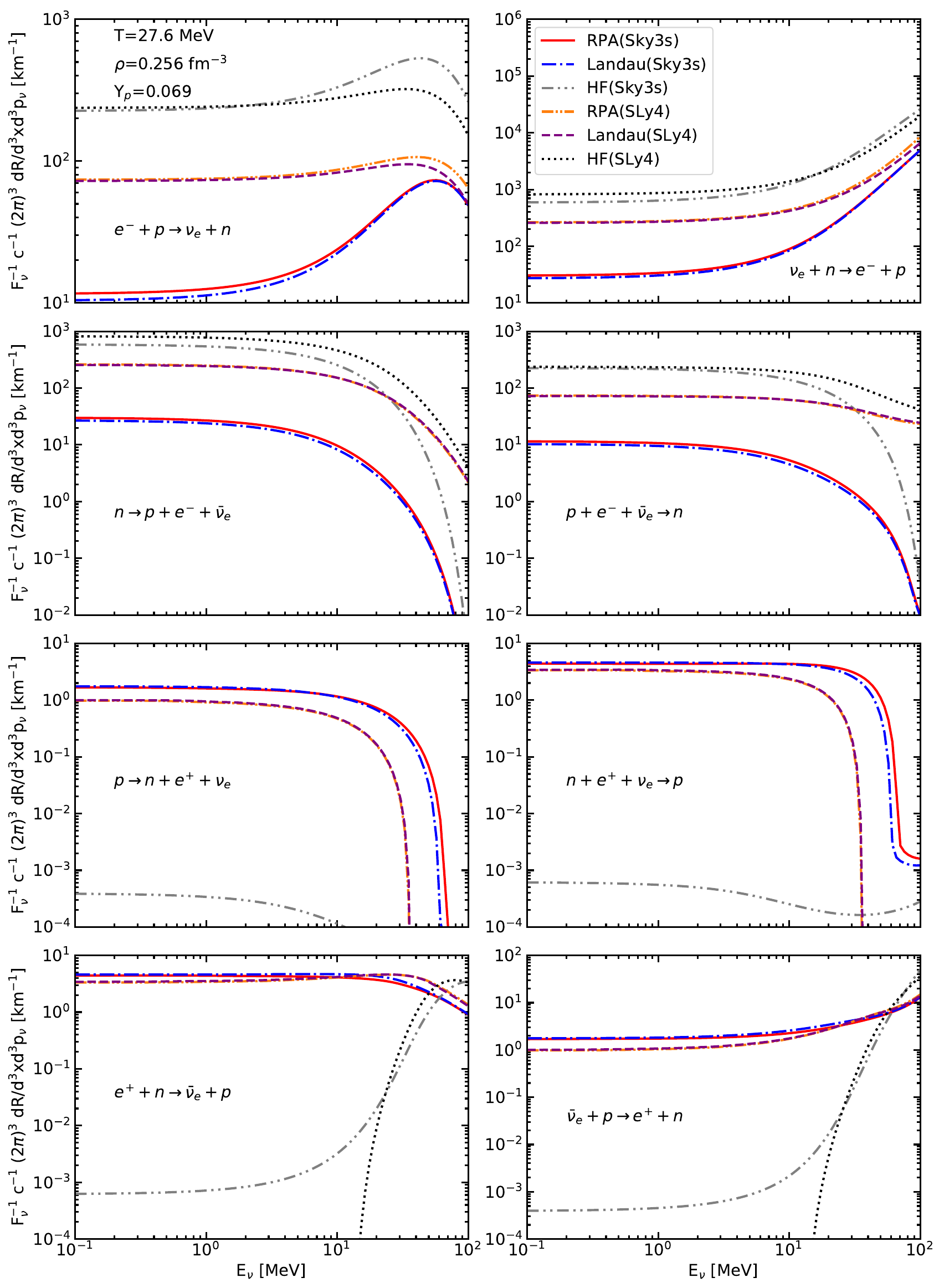} 
\caption{Similar to Fig. \ref{fig:neutrino-production-absorption-T19.5} but for $T=27.6$ MeV, $\rho=0.256$ fm$^{-3}$, and $Y_p=0.069$.}
\label{fig:neutrino-production-absorption-T27.6}
\end{figure*}

Figure \ref{fig:responses-n-p-T27.6} is similar to Fig. \ref{fig:responses-n-p-T19.5} but for a much higher density ($T=27.6$ MeV, $\rho=0.256$ fm$^{-3}$, $Y_p=0.069$).
The insets show again the entire peaks of the zero-sound modes.
The shape of the peak of the HF response for Sky3s is inverted compared to the one for SLy4 because the effective mass difference, i.e., $m_n^* - m_p^*$, has opposite signs for Sky3s and SLy4 (cf. Table \ref{table:effective-masses-central-potential}).
Thus the position of the HF response is more affected by the effective mass difference than by $\Delta U$.
Now there are sharp zero-sound modes in full RPA and Landau approximation in all four panels, and they concentrate almost the entire strength in an energy that can be quite far from the energy range of the HF response.
Notice that, instead of marking the positions of the zero-sound modes with arrows as in Ref. \cite{Duan2025}, the zero-sound modes are shown by Gaussian peaks with a finite width to present them more easily in the insets (when computing (anti)neutrino rates, we use smaller widths than in this figure).
An important difference is the position of the $S=1$ zero-sound mode in Sky3s and SLy4. The ph interaction in the $S=1$ channel in Sky3s is more repulsive than in SLy4, since the spin-dependent terms of Sky3s were adjusted to microscopic calculations of the $G_0$ and $G_0'$ Landau parameters \cite{Duan2025}. Therefore, with Sky3s, the $S=1$ zero-sound mode lies at a higher energy than with SLy4. We can see a difference in the peak positions between the full RPA and the Landau approximation, except in the $S=1$ channel computed with SLy4.
The positions of the zero-sound modes are important because they affect the threshold energies and thus the rates of various neutrino processes.

Now let us move on to the differential rates of (anti)neutrino production and absorption computed at different levels of approximation.
As we have seen above, in some cases, a large amount of the total strength of the response can be contained in zero-sound modes.
Hence, it is crucial to carefully treat these modes in the integral in Eq.~\eqref{eq:differential rates for all processes}.
In practice, as mentioned before, we subtract sharp (delta function) or very narrow peaks in the response functions and replace them by Gaussians containing the same strength as the subtracted peak.
Then the $q$ and $\omega$ integrations in Eq.~\eqref{eq:differential rates for all processes} are performed over smaller subintervals limited to the range of $\Im\Pi_0$ or of one of these Gaussians.

Figure \ref{fig:neutrino-production-absorption-T19.5} shows the differential rates of (anti)neutrino production (left panels) and absorption (right panels) as a function of the (anti)neutrino energy $E_{\nu}$ for interactions Sky3s and SLy4 at $T=19.5$ MeV, $\rho=0.00498$ fm$^{-3}$, $Y_p=0.0764$.
To facilitate the comparison with the results shown in Refs. \cite{Oertel2020,Pascal2022}, we multiply our differential rates $\frac{dR}{d^3 p_{\nu} d^3 x}$ by a factor of $(2\pi)^3/c$ ($c$ is the speed of light in units of km s$^{-1}$).
In fact, the results in the right panels are the absorption opacities, while the sum of the results in the left and corresponding right panels is the so-called absorption opacity corrected for stimulated absorption as shown in Refs. \cite{Oertel2020,Pascal2022}.

At the first glance, we can see that, compared with HF, the full RPA and the Landau approximation reduce the opacities in the processes with electrons (upper four panels).
This is consistent with previous findings in Refs. \cite{Burrows1999,Reddy1999,Dzhioev2018}.
But in the two processes with positrons (bottom panels), the full RPA and the Landau approximation enhance the antineutrino opacities since the zero-sound modes lower the energy threshold for these processes.
For the electron processes, the results computed with the full RPA and the Landau approximation are in reasonable agreement with each other, although there are small differences for Sky3s.
But for the positron processes, the full RPA and the Landau approximation give different threshold energies, not only with Sky3s but also with SLy4.
Notice that the two processes $p \rightarrow n + e^{+} + \nu_e$ and $n + e^{+} + \nu_e \rightarrow p$ cannot occur under this condition. 

Let us now compare the results for Sky3s and SLy4.
While the SLy4 results are compatible with those of Ref. \cite{Pascal2022}, the Sky3s results of the processes involving electrons are smaller by up to one order of magnitude, especially at lower energies, and for the positron processes it is the opposite.
Since these differences between the results computed with Sky3s and SLy4 appear already at the HF level, they probably come from the phase space allowed by the different nucleon effective masses, the different $\Delta U$, and the different $\Delta \mu$ as listed in Table \ref{table:effective-masses-central-potential}.

Figure \ref{fig:neutrino-production-absorption-T27.6} is similar to Fig. \ref{fig:neutrino-production-absorption-T19.5} but for $T=27.6$ MeV, $\rho=0.256$ fm$^{-3}$, $Y_p=0.069$.
Unlike the $T=19.5$ MeV case, the two processes $p \rightarrow n + e^{+} + \nu_e$ and $n + e^{+} + \nu_e \rightarrow p$ can occur under this condition.
Again, the results for the electron processes (upper four panels) computed with the full RPA and Landau approximation are smaller than those for HF.
This suppression and its evolution with energy are more drastic for Sky3s than for SLy4.
Conversely, the neutrino rates of positron processes (third row) computed with the full RPA and the Landau approximation are always larger than those computed with HF.
Finally, in the antineutrino processes with positrons (lowest two panels), the rates computed with the full RPA and the Landau approximation are larger (smaller) than those computed with HF for low-energy (higher-energy) antineutrinos.
This is also consistent with the finding of Ref. \cite{Dzhioev2018}.
It seems that the energies where the positron processes are kinematically allowed depend sensitively on the position of the zero-sound modes and therefore on the choice of the interaction.

We cannot reach a general conclusion about how the results computed with Sky3s differ from those computed with SLy4 based on only these two figures.
However, what is certain is that there are big differences between results computed with Sky3s and SLy4 because they predict different nucleon effective masses, different $\Delta U$, and different $\Delta \mu$.
Additional differences come from the spin-dependent terms, which determine the position of the zero-sound modes in the $S=1$ channels.
Since especially the effective masses and spin-dependent terms are better constrained in Sky3s than in SLy4, it may be worthwhile to repeat the simulations of Refs. \cite{Oertel2020,Pascal2022} with Sky3s. Although the dependence on the choice of the interaction is clearly much larger than the difference between the full RPA and the Landau approximation, we note that using rates obtained within the full RPA does not represent a big additional cost.

\begin{figure}[ht!]
\begin{center}
\includegraphics[scale=0.58]{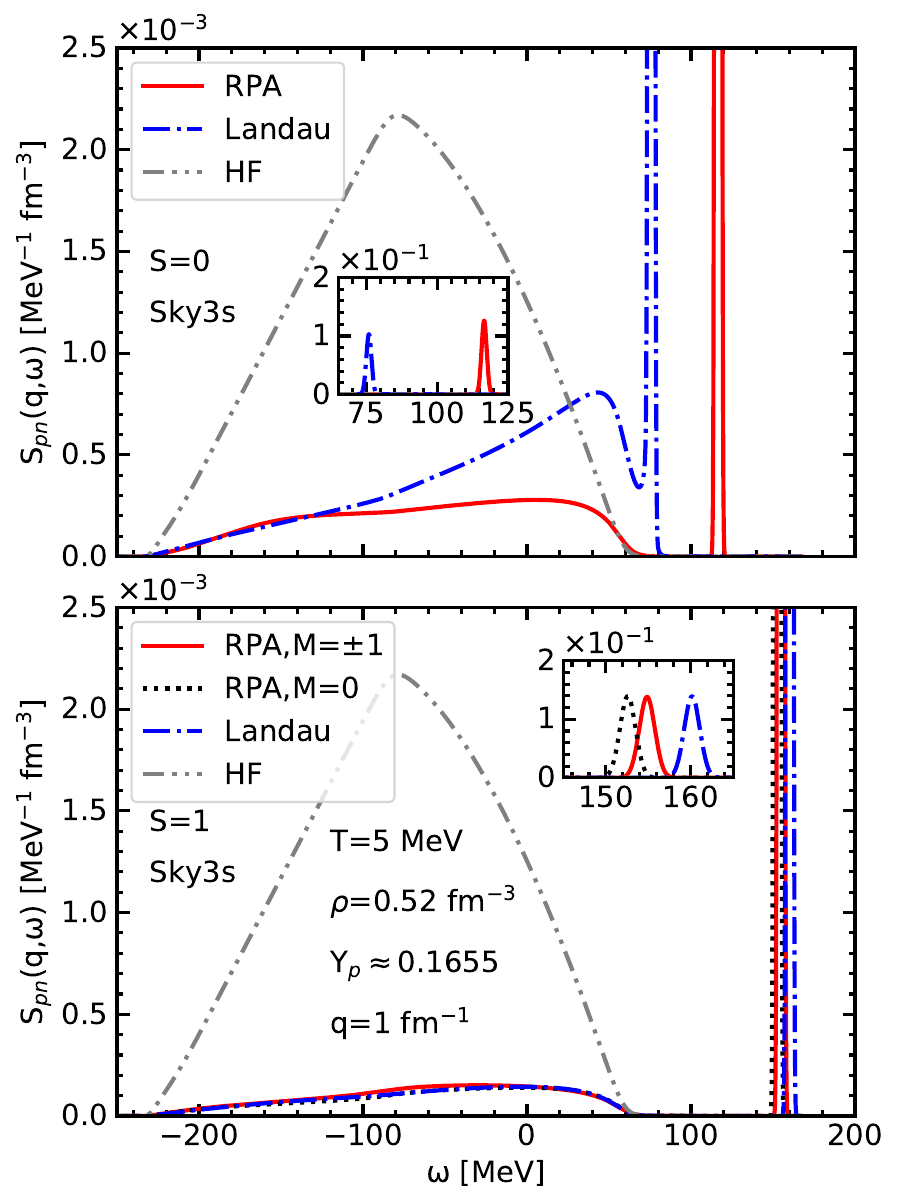} 
\caption{Response functions for Sky3s at $T=5$ MeV, $\rho=0.52$ fm$^{-3}$, $Y_p \approx 0.1655$ (corresponding to the standard $\beta$ equilibrium condition \eqref{eq:standard-beta-equilibrium} of $npe \mu$ matter) for $S=0$ (top) and $S=1$ (bottom).
  The momentum transfer is $q=1$ fm$^{-1}$.}
\label{fig:responses-n-p-T5}
\end{center}
\end{figure}

\begin{figure*}[ht!]
\centering
\includegraphics[scale=0.6]{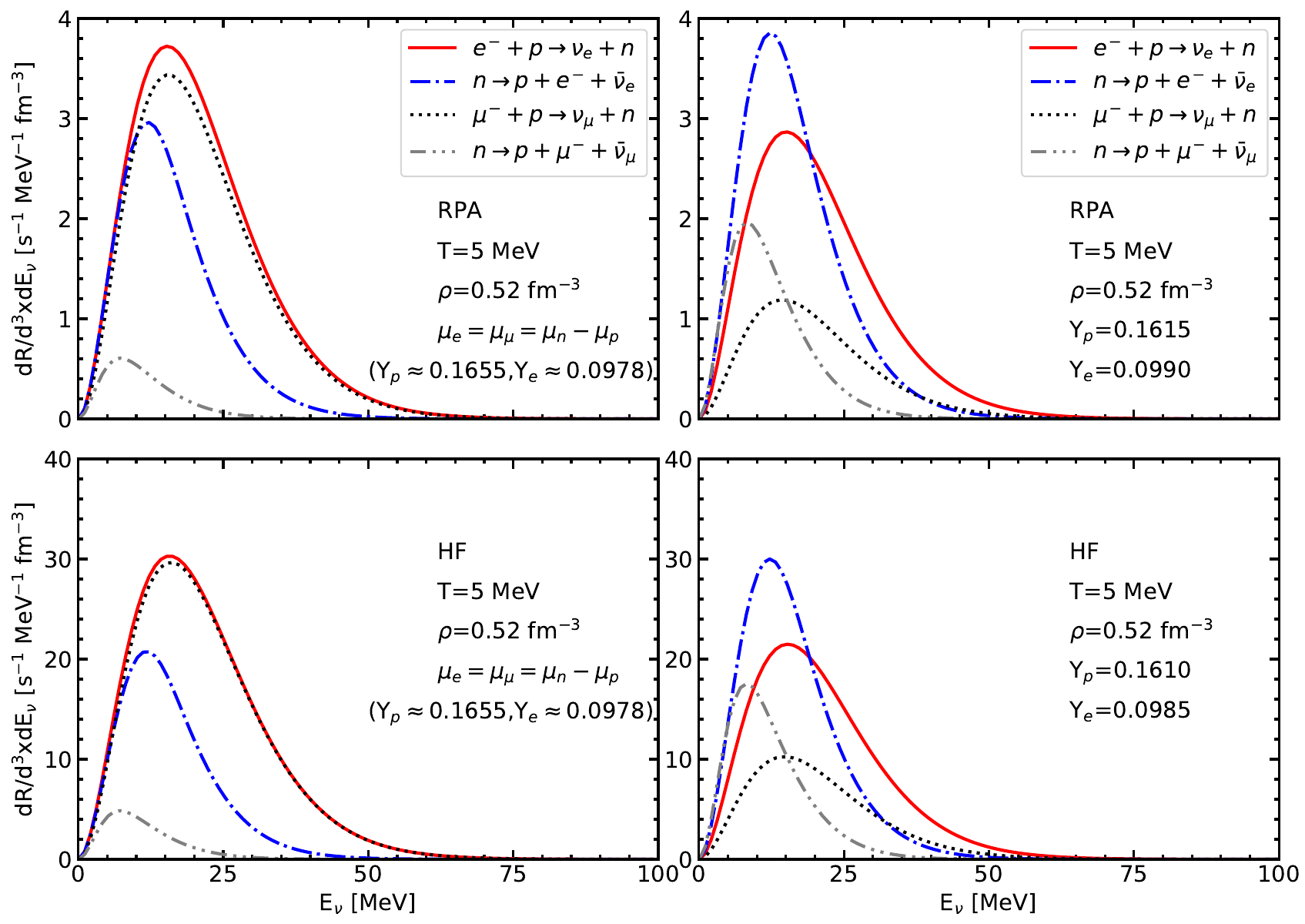} 
\caption{Differential production rates of (anti)neutrinos per energy and volume as a function of the energy $E_{\nu}$ for Sky3s in full RPA (top) and in HF approximation (bottom) at $T=5$ MeV and $\rho=0.52$ fm$^{-3}$. Left: under the standard $\beta$ equilibrium condition \eqref{eq:standard-beta-equilibrium}; right: under the modified $\beta$ equilibrium condition \eqref{eq:modified-beta-equilibrium}.}
\label{fig:dUrca-beta-equilibrium}
\end{figure*}

\subsection{Direct Urca neutrino emission from matter in beta equilibrium}
\label{subsec:dUrca}

The initial internal temperature of a newborn neutron star is of the order of $10^{11} - 10^{12}$ K (10 MeV-100 MeV) \cite{HaenselPotekhinYakovlev,DegenaarSuleimanov}, and gradually cools down mainly via neutrino emissions \cite{Yakovlev2004}. The dominant neutrino processes are known as Urca processes, including direct Urca and modified Urca processes. The direct Urca processes are the most efficient ones, and they occur when the energy and momentum can be simultaneously conserved \cite{Lattimer1991}:
\begin{equation}\label{eq:dUrca-electron}
e^{-} +p \rightarrow n + \nu_e, \quad n \rightarrow p + e^{-} + \bar{\nu}_e.
\end{equation}
In addition to electrons, muons are expected to appear at sufficiently high densities, when $\mu_e > m_{\mu}$. Then also direct Urca processes with muons can occur \cite{Lattimer1991}:
\begin{equation}\label{eq:dUrca-muon}
\mu^{-} + p \rightarrow n + \nu_{\mu}, \quad n \rightarrow p + \mu^{-} + \bar{\nu}_{\mu}.
\end{equation}

In this section, we study the direct Urca neutrino emission from hot neutron star matter in $\beta$ equilibrium.
At low temperatures, neutron stars described with SLy4 cannot cool down via the direct Urca processes \cite{Chabanat1997}, while with Sky3s the threshold density for direct Urca processes is $\approx 0.47$ fm$^{-3}$ \cite{Duan2024}.
Therefore, we present only the results computed with Sky3s.
Since our theory does not include modified Urca processes and to avoid spurious numerical effects, we choose a sufficiently high temperature and a number density that is slightly higher than the threshold value of the direct Urca processes, namely, $T=5$ MeV and $\rho=0.52$ fm$^{-3}$.
The corresponding proton fraction is $Y_p \approx 0.1655$ under the standard $\beta$ equilibrium of $npe \mu$ matter defined by
\begin{equation}\label{eq:standard-beta-equilibrium}
    \mu_n-\mu_p = \mu_e = \mu_\mu\,.
\end{equation}
Such conditions are interesting in the context of neutron-star mergers, where matter in $\beta$ equilibrium is heated \cite{Alford2018}.

As in Sec. \ref{subsec:results-protoneutron-star}, we show first the response functions in Figure \ref{fig:responses-n-p-T5}.
Since the momentum transfer in direct Urca processes is of the order of $k_{F,n}-k_{F,p}$, we choose $q=1$ fm$^{-1}$.
In the top and bottom panels, the $S=0$ and $S=1$ responses are shown, respectively.
Unlike the cases shown in Figs. \ref{fig:responses-n-p-T19.5} and \ref{fig:responses-n-p-T27.6}, there is a visible difference between the ($S=1, M=\pm1$) and ($S=1,M=0$) channels in the present case. In the $S=0$ channel, the difference between results computed with the full RPA and Landau approximation is quite large:
The Landau approximation underestimates the ph interaction and therefore the zero-sound mode lies much closer to the ph continuum and absorbs less strength than in the full RPA.

Assuming that all produced neutrinos immediately leave the star (no Pauli blocking: $n^\nu_{\pv_{\nu}}=0$), and using $d^3 p_{\nu} = 4\pi E_{\nu}^2 dE_{\nu}$, the differential production rates of (anti)neutrinos per energy and volume can be obtained as (they do not depend on the direction of $\pv_{\nu}$)
\begin{equation}\label{eq:neutrino-rate-per-energy-volume}
\frac{dR}{dE_{\nu} d^3x} = 4\pi E_{\nu}^2 \frac{dR}{d^3 p_{\nu} d^3 x}.
\end{equation}
Therefore, we show in this section the differential production rates of (anti)neutrinos per energy and volume, and the total production rates per volume, $dR/d^3x$, can be directly read off from their integrals over $E_\nu$.
The left panels of Fig. \ref{fig:dUrca-beta-equilibrium} show the results computed with the full RPA (top) and with HF (bottom) under the same standard $\beta$ equilibrium condition as in Fig. \ref{fig:responses-n-p-T5}.
We can see that, for both RPA and HF, the total electron neutrino rates are not equal to the total electron antineutrino rates. This is also true for muon neutrinos and antineutrinos.

But instead of considering the neutrinos, which leave the star, let us now focus on what this implies for the composition of matter. Namely, the total electron capture rate (integral of the red solid line) is larger than the electron production rate (blue dash-dotted line), and the muon capture rate (black dotted line) is larger than the muon production rate (gray dash-dot-dot line). Hence, the composition obtained from the standard $\beta$ equilibrium condition \eqref{eq:standard-beta-equilibrium} does not correspond to a stationary one:   $Y_e$, $Y_{\mu}$, and $Y_p$ increase with time while $Y_n = 1-Y_p$ decreases.
It means that the standard $\beta$ equilibrium condition is not true any more at finite temperature.
This finding is consistent with the point of view in Ref. \cite{Alford2018}.

To find the true $\beta$ equilibrium condition at finite temperature (which we will call the modified $\beta$ equilibrium), we should find the composition that results in equal total neutrino and antineutrino rates in both electron and muon cases, i.e., solve the two equations
  \begin{align}\label{eq:modified-beta-equilibrium}
    R_{e^-+p\to \nu_e+n} &= R_{n\to p+e^-+\nubar_e}\,,\nonumber \\
    R_{\mu^-+p\to \nu_{\mu}+n} &= R_{n\to p+\mu^-+\nubar_{\mu}}\,,
  \end{align}
  for the two unknown fractions $Y_e$ and $Y_{\mu}$, keeping $Y_p=Y_e+Y_{\mu}$ (charge neutrality).
We find that under the modified $\beta$ equilibrium condition \eqref{eq:modified-beta-equilibrium}, the proton and electron fractions get slightly modified to $Y_p \approx 0.1615$, $Y_e \approx 0.0990$ for the RPA case and $Y_p \approx 0.1610$, $Y_e \approx 0.0985$ for the HF case.

The effective masses, $\Delta U$, $\Delta \mu$, chemical potentials of electrons and muons under these conditions can be found in Table \ref{table:effective-masses-central-potential}.
An interesting observation is that the modified $\beta$ equilibrium conditions for RPA and HF are slightly different.
We can see that the standard $\beta$ equilibrium condition \eqref{eq:standard-beta-equilibrium} is violated by 2 MeV (2.7 MeV) for electrons and by 7.5 MeV (7.8 MeV) for muons in the RPA (HF) case.
This is much less dramatic than the violations found in Ref. \cite{Alford2018} for similar densities and temperatures (cf. lower panel of Fig. 3 in Ref. \cite{Alford2018}).
However, the two situations are not comparable because $\rho=0.52$ fm$^{-3}$ lies above the direct Urca threshold for Sky3s ($\approx 3\rho_0$) while it lies below the direct Urca threshold for the APR equation of state ($\approx 5\rho_0$) used in Ref. \cite{Alford2018}.

The (anti)neutrino spectra computed at the modified $\beta$ equilibrium are shown in the right panels of Fig. \ref{fig:dUrca-beta-equilibrium}. We can see that the spectra of neutrinos are different from those of antineutrinos, although the total neutrino rates are equal to the total antineutrino rates. The neutrinos have a higher average energy, i.e., they contribute more to the cooling, than the antineutrinos.

\section{Conclusion}
\label{sec:conclusion}
This work aims to study charged current neutrino processes in hot nuclear matter with the recent extended Skyrme parametrization Sky3s, whose effective masses and spin dependent terms were adjusted to results of microscopic BHF calculations.
First, we recalled the relation between the differential rates of (anti)neutrino production (absorption) and the charged-current response functions.
Then we derived the full RPA response functions for extended Skyrme forces using an alternative (but equivalent) method to the one of \cite{Davesne2019}. The more commonly used Landau approximation can be easily obtained as a simplified version of the RPA.

Then, we showed the results for response functions computed with Sky3s and those computed with the widely used SLy4 parametrization for comparison.
The resulting neutrino production and absorption in proto-neutron star and supernova matter, and direct Urca neutrino emission from $\beta$ stable neutron star matter were studied.

Compared to the HF results, both the full RPA and the Landau approximation strongly reduce the (anti)neutrino rates in the processes with electrons, which is consistent with the findings in the previous literature \cite{Reddy1999,Dzhioev2018,Burrows1999}. However, this is not always true in the processes with positrons.
Although there are in general some differences between the results computed with the full RPA and with the Landau approximation, they agree more or less with each other and the differences will probably not be significant for astrophysical simulations.
Nevertheless, in the case of Skyrme functionals (unlike other interactions), using the full RPA instead of the Landau approximation does not represent a big additional effort, and hence there is no good reason for not using it.

The results computed with Sky3s differ from those computed previously with SLy4 because of the different nucleon effective masses, mean fields, and consequently also different chemical potentials.
We recall that the SLy4 effective masses decrease so strongly with density that the neutron Fermi velocity exceeds the speed of light at $\rho\gtrsim 2\rho_0$ \cite{Duan2023}, while the Sky3s effective masses were adjusted to microscopic calculations where such a pathological behavior does not occur \cite{Duan2024}. This changes the available phase space already at the HF level and can lead to dramatic differences in the rates of one order of magnitude. Beyond the HF level, additional differences between Sky3s and SLy4 may come from ph interaction in the $S=1$ channel, which is more repulsive in Sky3s at high densities since it was constrained by microscopic calculations \cite{Duan2025}.

We also studied the rates of electron and muon weak interaction processes in $\beta$ stable $npe\mu$ matter with Sky3s above the direct Urca threshold.
As pointed out in Ref. \cite{Alford2018}, the standard $\beta$ equilibrium condition does not lead to a stationary composition of matter. We found that by slightly changing the electron, muon, and proton fractions, a stationary composition of matter can be obtained.
In this modified $\beta$ equilibrium, the spectra of neutrinos differ from those of antineutrinos, although the total neutrino rates are equal to the total antineutrino rates.
Quantitatively, we find a smaller deviation from the standard $\beta$ equilibrium than predicted in Ref. \cite{Alford2018}, probably because of the much lower direct Urca threshold in Sky3s.

Recently, RPA calculations of neutrino absorption rates in $\beta$
equilibrium using interactions from chiral EFT were presented in
Ref. \cite{Shin2024}, and we performed calculations for the same
densities and temperatures in order to compare with those results. To
avoid lengthiness, we do not show them here. While there is good
agreement at very low density ($\rho=0.002$ fm$^{-3}$), there are
significant differences between Sky3s and chiral results in the case
$\rho = 0.02$ fm$^{-3}$, $T=8$ MeV (but they are smaller than the
differences between SLy4 and chiral results). We suspect that the main
reason for these differences is the mean field shift $U_n-U_p$, on
which the proton fraction and the neutrino rates depend very
sensitively. For $\rho = 0.02$ fm$^{-3}$, the HF shift in
Ref. \cite{Shin2024} ($\Delta U = 10$ MeV) is much weaker than in our
Sky3s calculation ($\Delta U = 17.7$ MeV). However, to reach converged
results with chiral interactions, it may be necessary to include
corrections beyond HF to the single-particle properties and
particle-hole interactions \cite{Alp2025}. This important question is
currently investigated by several groups.

In conclusion, the predictions of neutrino rates depend strongly on the approximation scheme (HF or RPA) and on the used nuclear interaction. Differences can be one order of magnitude or even more and they may therefore be relevant for astrophysical simulations.
Hence, it is important to use the best available method for astrophysical calculations, which in the present case is RPA and not HF.
Furthermore, interactions such as SLy4, which present pathologies and unconstrained spin dependent terms already at relatively moderate densities, cannot be expected to give reliable predictions of the neutrino rates.
Therefore, we plan to compute tables of neutrino rates for astrophysical simulations also with other interactions, such as Sky3s which has better constrained effective masses and spin dependent terms.

Furthermore, within the present theoretical framework, we can only treat processes with one incoming and one outgoing nucleon. However, for $\beta$ stable matter at low temperatures, the direct Urca process is not possible below some threshold density, and with some interactions not at all. In this case, modified Urca processes involving a second nucleon become important. Recently, it was suggested that these processes can be approximately treated like single-nucleon processes by giving a finite width to the nucleon \cite{Sedrakian2024}. Another important effect not included here is pairing. Below the transition temperature towards the superfluid state, the pair-breaking and formation process is the most efficient neutrino process when direct Urca processes are not allowed \cite{Flowers1976}.

It is also worth noticing that relativistic effects may become
increasingly important at very high densities. The Fermi interaction
that we used here is only the leading order in $p/m_N$ (where $p$ and
$m_N$ are the nucleon momentum and mass). An important relativistic
correction is expected from weak magnetism
\cite{Horowitz2002,Leinson2002,Roberts2017} because it is enhanced by
the large isovector anomalous magnetic moment of the
nucleons. Calculations in the literature found that it can increase
the total neutrino emissivity by up to 50\% \cite{Leinson2002}. Apart
from that, relativistic kinematics may also become relevant
\cite{Roberts2017,Fischer2020a,Guo2020,Sugiura2022,Sokolowski2026}.
These questions are left to future studies.

\acknowledgments We thank M. Oertel for useful discussions and for sending us the results of Ref. \cite{Pascal2022} that we used to check our code.
\section*{Data availability}
The numerical results shown in Figs. 1--6 are available from Ref. \cite{Duan2026data}.

\appendix
\section{Expressions of $A_{ik}$}\label{sec:expression of Aik}
The non-vanishing matrix elements $A_{ik}$ are given below:
\begin{align}\label{eq:matrix elements of A}
 A_{1,1} &= v_{1}^{0} \Pi_{0} + v_{2}^{0} \Pi_{2}\,,  & A_{1,28} &= v_{1}^{0} \Pi_{2} + v_{2}^{0} \Pi_{4}\,,\nonumber\\
 A_{1,30} &= 2 v_{8}^{0} q^{2}  \Pi_{2T}\,,          & A_{1,35} &= v_{1}^{0} \Pi_{A} + v_{2}^{0} \Pi_{B}\,, \notag \\
 A_{2,2} &= v_{1}^{0} \Pi_{0} + v_{2}^{0} \Pi_{2}\,,  & A_{2,10} &= v_{1}^{0} \Pi_{2} + v_{2}^{0} \Pi_{4}\,, \nonumber\\
 A_{2,16} &= 2 v_{8}^{0} q^{2} \Pi_{2T}\,,           & A_{2,36} &= v_{1}^{0} \Pi_{A} + v_{2}^{0} \Pi_{B}\,, \notag \\
 A_{3,3} &= v_{3}^{0} \Pi_{2T}\,,                   & A_{3,9}  &= v_{8}^{0} q^{2} \Pi_{0}\,,\nonumber\\
 A_{3,34} &= v_{8}^{0} q^{2} \Pi_{2}\,,              & A_{3,42} &= v_{8}^{0} q^{2} \Pi_{A}\,, \notag \\
 A_{4,4} &=v_{4}^{0} \Pi_{0} + v_{5}^{0} \Pi_{2}\,,   & A_{4,29} &= v_{4}^{0} \Pi_{2} + v_{5}^{0} \Pi_{4}\,,\nonumber\\
 A_{4,31} &= v_{8}^{0}  q^{2} \Pi_{2T}\,,            & A_{4,37} &= v_{4}^{0} \Pi_{A} + v_{5}^{0} \Pi_{B}\,, \notag \\
 A_{5,5} &= v_{4}^{0} \Pi_{0} + v_{5}^{0} \Pi_{2}\,, & A_{5,12} &= v_{4}^{0} \Pi_{2} + v_{5}^{0} \Pi_{4}\,,\nonumber\\
 A_{5,17} &= v_{8}^{0} q^{2} \Pi_{2T}\,,            & A_{5,38} &= v_{4}^{0} \Pi_{A} + v_{5}^{0} \Pi_{B}\,, \notag \\
 A_{6,6} &= v_{6}^{0} \Pi_{2T}\,,                   & A_{7,7} &= v_{4}^{0} \Pi_{0} + v_{5}^{0} \Pi_{2}\,,\nonumber\\ 
 A_{7,31} &= -v_{8}^{0} \Pi_{2T}\,,                 & A_{7,32} &= v_{4}^{0} \Pi_{2} + v_{5}^{0} \Pi_{4}\,, \notag \\
 A_{7,39} &= v_{4}^{0} \Pi_{A} + v_{5}^{0} \Pi_{B}\,, & A_{8,6} &= v_{8}^{0} \Pi_{2T}\,,\nonumber\\
 A_{8,8} &= v_{1}^{0} \Pi_{0} + v_{2}^{0} \Pi_{2}\,,  & A_{8,18} &= 2  v_{8}^{0} q^{2} \Pi_{2T}\,,  \notag \\
 A_{8,33} &= v_{1}^{0} \Pi_{2} + v_{2}^{0} \Pi_{4}\,, & A_{8,41} &= v_{1}^{0} \Pi_{A} + v_{2}^{0} \Pi_{B}\,,\nonumber\\ 
 A_{9,3} &= v_{8}^{0} \Pi_{2T} \,,                  & A_{9,9} &= v_{4}^{0} \Pi_{0} + v_{5}^{0} \Pi_{2} \notag \\
 A_{9,34} &= v_{4}^{0} \Pi_{2} + v_{5}^{0} \Pi_{4}\,, & A_{9,42} &=  v_{4}^{0} \Pi_{A} + v_{5}^{0} \Pi_{B}\,,\nonumber\\
 A_{10,2} &= v_{2}^{0} \Pi_{0} \,,                 & A_{10,10} &=  v_{2}^{0} \Pi_{2}\,, \notag \\
 A_{10,36} &=  v_{2}^{0} \Pi_{A}\,,                 & A_{11,3} &= \frac{v_{3}^{0} (\Pi_{2L} - \Pi_{2T})}{q^{2}}\,, \nonumber\\
 A_{11,9} &=  -v_{8}^{0} \Pi_{0} \,,                & A_{11,11} &= v_{3}^{0} \Pi_{2L} \notag \\
 A_{11,19} &= \frac{1}{q^{2}} v_{3}^{0} \Pi_{A} \,,  & A_{11,20} &= \frac{1}{q^{2}} v_{3}^{0} \Pi_{B}\,,\nonumber\\
 A_{11,34} &= -v_{8}^{0} \Pi_{2} \,,                & A_{11,42} &= - v_{8}^{0} \Pi_{A}\,, \notag \\
 A_{12,5} &= v_{5}^{0} \Pi_{0} \,,                  & A_{12,12} &= v_{5}^{0} \Pi_{2}\,,\nonumber\\
 A_{12,38} &=  v_{5}^{0} \Pi_{A} \,,                & A_{13,6} &= \frac{v_{6}^{0} (\Pi_{2L} - \Pi_{2T})}{q^{2}}\,, \notag \\
 A_{13,13} &= v_{6}^{0} \Pi_{2L} \,,                & A_{13,21} &= \frac{1}{q^{2}} v_{6}^{0} \Pi_{A} \,,\nonumber\\
 A_{13,22} &= \frac{1}{q^{2}} v_{6}^{0} \Pi_{B} \,, & A_{14,14} &= v_{4}^{0} \Pi_{0} + v_{5}^{0} \Pi_{2}\,, \notag \\
 A_{14,15} &= v_{4}^{0} \Pi_{2} + v_{5}^{0} \Pi_{4}\,, & A_{14,17} &= -v_{8}^{0} \Pi_{2T}\,, \nonumber\\
 A_{14,40} &= v_{4}^{0} \Pi_{A} + v_{5}^{0} \Pi_{B} \,, & A_{15,14} &= v_{5}^{0} \Pi_{0}\,, \notag \\
 A_{15,15} &= v_{5}^{0} \Pi_{2} \,,                 & A_{15,40} &=  v_{5}^{0} \Pi_{A}\,,\nonumber\\ 
 A_{16,2} &= v_{8}^{0} \Pi_{0} \,,                  & A_{16,10} &=  v_{8}^{0} \Pi_{2}\,, \notag \\
 A_{16,16} &= v_{6}^{0} \Pi_{2T} \,,                & A_{16,36} &=  v_{8}^{0} \Pi_{A}\,, \nonumber\\
 A_{17,5} &= v_{8}^{0} \Pi_{0} \,,                  & A_{17,12} &= v_{8}^{0} \Pi_{2}\,,\notag \\
 A_{17,17} &=  v_{3}^{0} \Pi_{2T} \,,               & A_{17,38} &=  v_{8}^{0} \Pi_{A} \,,\nonumber\\
 A_{18,8} &=  v_{8}^{0} \Pi_{0} \,,                 & A_{18,18} &= v_{6}^{0} \Pi_{2T}\,,  \notag \\
 A_{18,33} &=  v_{8}^{0} \Pi_{2} \,,                & A_{18,41} &=  v_{8}^{0} \Pi_{A}\,, \nonumber\\
 A_{19,3} &= \frac{v_{1}^{0}\Pi_{A}+v_{2}^{0}\Pi_{B}}{q^2}\,,& A_{19,11} &= v_{1}^{0} \Pi_{A} + v_{2}^{0} \Pi_{B}\,, \notag \\
 A_{19,19} &= v_{1}^{0} \Pi_{0} + v_{2}^{0} \Pi_{2}\,,& A_{19,20} &= v_{1}^{0} \Pi_{2} + v_{2}^{0} \Pi_{4}\,,\nonumber\\
 A_{19,26} &= 2 v_{8}^{0} q^{2} \Pi_{2T} \,,        & A_{20,3} &= \frac{1}{q^{2}} v_{2}^{0} \Pi_{A}\,, \notag \\
 A_{20,11} &=  v_{2}^{0} \Pi_{A} \,,               & A_{20,19} &= v_{2}^{0} \Pi_{0} \,,\nonumber\\
 A_{20,20} &= v_{2}^{0} \Pi_{2} \,,                & A_{21,6} &= \frac{v_{4}^{0} \Pi_{A} + v_{5}^{0} \Pi_{B}}{q^2}\,, \notag \\
 A_{21,13} &= v_{4}^{0}\Pi_{A} + v_{5}^{0}\Pi_{B}\,,  & A_{21,21} &= v_{4}^{0} \Pi_{0} + v_{5}^{0} \Pi_{2}\,,\nonumber\\
 A_{21,22} &= v_{4}^{0}\Pi_{2} + v_{5}^{0}\Pi_{4} \,, & A_{21,27} &= v_{8}^{0} q^{2} \Pi_{2T}\,, \notag \\
 A_{22,6} &= \frac{1}{q^{2}} v_{5}^{0} \Pi_{A} \,,   & A_{22,13} &= v_{5}^{0} \Pi_{A}\,, \nonumber\\
 A_{22,21} &= v_{5}^{0} \Pi_{0} \,,                 & A_{22,22} &= v_{5}^{0} \Pi_{2}\,,  \notag \\
 A_{23,23} &= v_{4}^{0}\Pi_{0} + v_{5}^{0}\Pi_{2} \,, & A_{23,24} &= v_{4}^{0} \Pi_{2} + v_{5}^{0} \Pi_{4} \,,\nonumber\\
 A_{23,25} &= v_{4}^{0}\Pi_{A} + v_{5}^{0}\Pi_{B} \,, & A_{23,27} &= -v_{8}^{0} \Pi_{2T}\,, \notag \\
 A_{24,23} &= v_{5}^{0} \Pi_{0} \,,                 & A_{24,24} &=  v_{5}^{0} \Pi_{2}\,, \nonumber\\
 A_{24,25} &= v_{5}^{0} \Pi_{A} \,,                 & A_{25,23} &= \frac{1}{q^{2}} v_{6}^{0} \Pi_{A}\,, \notag \\
 A_{25,24} &= \frac{1}{q^{2}} v_{6}^{0} \Pi_{B} \,, & A_{25,25} &= v_{6}^{0} \Pi_{2L}  \,,\nonumber\\
 A_{26,3} &= \frac{1}{q^{2}} v_{8}^{0} \Pi_{A} \,,  & A_{26,11} &=  v_{8}^{0} \Pi_{A}\,, \notag \\
 A_{26,19} &= v_{8}^{0} \Pi_{0} \,,                & A_{26,20} &= v_{8}^{0} \Pi_{2}\,, \nonumber\\
 A_{26,26} &= v_{6}^{0} \Pi_{2T} \,,                & A_{27,6} &= \frac{1}{q^{2}} v_{8}^{0} \Pi_{A}\,, \notag \\
 A_{27,13} &=  v_{8}^{0} \Pi_{A} \,,                & A_{27,21} &= v_{8}^{0} \Pi_{0} \,,\nonumber\\
 A_{27,22} &= v_{8}^{0} \Pi_{2} \,,                 & A_{27,27} &= v_{3}^{0} \Pi_{2T}\,, \notag \\
 A_{28,1} &= v_{2}^{0} \Pi_{0} \,,                 & A_{28,28} &= v_{2}^{0} \Pi_{2} \,,\nonumber\\
 A_{28,35} &=  v_{2}^{0} \Pi_{A} \,,                &A_{29,4} &= v_{5}^{0} \Pi_{0}\,, \notag \\
 A_{29,29} &= v_{5}^{0} \Pi_{2} \,,                 & A_{29,37} &= v_{5}^{0} \Pi_{A}\,,\nonumber\\
 A_{30,1} &= v_{8}^{0} \Pi_{0} \,,                  & A_{30,28} &= v_{8}^{0} \Pi_{2}\,,\notag\\
 A_{30,30} &= v_{6}^{0} \Pi_{2T} \,,                & A_{30,35} &=  v_{8}^{0} \Pi_{A}\,,\nonumber\\ 
 A_{31,4} &= v_{8}^{0} \Pi_{0} \,,                  & A_{31,29} &= v_{8}^{0} \Pi_{2} \notag \\
 A_{31,31} &= v_{3}^{0} \Pi_{2T} \,,                & A_{31,37} &=  v_{8}^{0} \Pi_{A}\,,\nonumber\\
 A_{32,7} &= v_{5}^{0} \Pi_{0} \,,                  & A_{32,32} &= v_{5}^{0} \Pi_{2}\,, \notag \\
 A_{32,39} &= v_{5}^{0} \Pi_{A} \,,                 & A_{33,8} &= v_{2}^{0} \Pi_{0} \,,\nonumber\\
 A_{33,33} &= v_{2}^{0} \Pi_{2} \,,                & A_{33,41} &=  v_{2}^{0} \Pi_{A}\,, \notag \\
 A_{34,9} &= v_{5}^{0} \Pi_{0} \,,                 & A_{34,34} &= v_{5}^{0} \Pi_{2} \,,\nonumber\\
 A_{34,42} &=  v_{5}^{0} \Pi_{A} \,,               & A_{35,1} &= \frac{1}{q^{2}} v_{3}^{0} \Pi_{A} \,, \notag \\
 A_{35,28} &= \frac{1}{q^{2}} v_{3}^{0} \Pi_{B} \,, & A_{35,35} &= v_{3}^{0} \Pi_{2L} \,,\nonumber\\
 A_{36,2} &= \frac{1}{q^{2}} v_{3}^{0} \Pi_{A} \,,  & A_{36,10} &= \frac{1}{q^{2}} v_{3}^{0} \Pi_{B} \,, \notag \\
 A_{36,36} &= v_{3}^{0} \Pi_{2L} \,,                & A_{37,4} &= \frac{1}{q^{2}} v_{6}^{0} \Pi_{A} \,,\nonumber\\
 A_{37,29} &= \frac{1}{q^{2}} v_{6}^{0} \Pi_{B} \,, & A_{37,37} &= v_{6}^{0} \Pi_{2L} \,, \notag \\
 A_{38,5} &= \frac{1}{q^{2}} v_{6}^{0} \Pi_{A} \,,  & A_{38,12} &= \frac{1}{q^{2}} v_{6}^{0} \Pi_{B} \,, \nonumber\\
 A_{38,38} &= v_{6}^{0} \Pi_{2L} \,,                & A_{39,7} &= \frac{1}{q^{2}} v_{6}^{0} \Pi_{A} \,, \notag \\
 A_{39,32} &= \frac{1}{q^{2}} v_{6}^{0} \Pi_{B} \,, & A_{39,39} &= v_{6}^{0} \Pi_{2L} \,,\nonumber\\
 A_{40,14} &= \frac{1}{q^{2}} v_{6}^{0} \Pi_{A} \,, & A_{40,15} &= \frac{1}{q^{2}} v_{6}^{0} \Pi_{B} \,, \notag \\
 A_{40,40} &= v_{6}^{0} \Pi_{2L} \,,                & A_{41,8} &= \frac{1}{q^{2}} v_{3}^{0} \Pi_{A} \,,\nonumber\\
 A_{41,33} &= \frac{1}{q^{2}} v_{3}^{0} \Pi_{B} \,, & A_{41,41} &= v_{3}^{0} \Pi_{2L} \,, \notag \\
 A_{42,9} &= \frac{1}{q^{2}} v_{6}^{0} \Pi_{A} \,, & A_{42,34} &= \frac{1}{q^{2}} v_{6}^{0} \Pi_{B}\,,\nonumber\\
 A_{42,42} &= v_{6}^{0} \Pi_{2L}\,.
\end{align}

\bibliography{references}

\end{document}